\documentclass[12pt]{article} 

\usepackage[utf8]{inputenc}
\usepackage[T1]{fontenc} 
\usepackage[english]{babel}           

\usepackage[pdftex]{graphicx}
\usepackage{epsfig}
\usepackage{graphicx}
\usepackage{comment}
\usepackage{latexsym}
\usepackage{hyperref}
\usepackage{amsmath}
\usepackage{mathrsfs}
\usepackage[usenames, dvipsnames]{color}
\usepackage{amsbsy}
\usepackage{amssymb}
\usepackage{amsthm}
\usepackage{amsfonts}
\usepackage{cite}
\usepackage{enumitem}
\usepackage{xcolor}
\usepackage{color}
\usepackage{mathtools}
\usepackage{caption} 
\usepackage{subcaption} 

\usepackage{diagbox}

\newcommand{\be}[0]{\begin{equation}}
\newcommand{\ee}[0]{\end{equation}}
\newcommand{\dis}{\displaystyle}
\renewcommand{\thefootnote}{\fnsymbol{footnote}}

\newcommand{\R}{\mathbb{R}}
\newcommand{\Z}{\mathbb{Z}}

\renewcommand{\O}{{\cal O}}

\newcommand{\ie}{{\em i.e.} }
\newcommand{\eg}{{\em e.g.} }
\newcommand{\via}{{\it via} }

\newcommand{\where}{\mbox{where}}

\renewcommand{\and}{\mbox{and}}

\newcommand{\espD}{\phantom{\!\!\underset{\displaystyle |}{\cdot}}}

\newcommand{\bm}{\boldmath} 

\newcommand{\F}{{\cal F}}
\newcommand{\N}{{\cal N}}

\newcommand{\K}{{\cal K}}

\renewcommand{\S}{{\cal S}}

\newcommand{\cR}{{\cal R}}
\newcommand{\Hc}{{\cal H}}

\newcommand{\Ms}{M_{\rm s}}

\newcommand{\nF}{n_{\rm F}}
\newcommand{\nB}{n_{\rm B}}
\newcommand{\nFt}{\tilde n_{\rm F}}
\newcommand{\nBt}{\tilde n_{\rm B}}

\newcommand\dd{\text{d}}

\def\jac(#1,#2){%
\begin{bsmallmatrix}
#1\cr 
#2\cr
\end{bsmallmatrix}}

\catcode`\@=11
\def\marginnote#1{}
\newcount\hour
\newcount\minute
\newtoks\amorpm
\hour=\time\divide\hour by60 \minute=\time{\multiply\hour by60
\global\advance\minute by-\hour}
\edef\standardtime{{\ifnum\hour<12 \global\amorpm={am}%
        \else\global\amorpm={pm}\advance\hour by-12 \fi
        \ifnum\hour=0 \hour=12 \fi
        \number\hour:\ifnum\minute<10 0\fi\number\minute\the\amorpm}}
\edef\militarytime{\number\hour:\ifnum\minute<10 0\fi\number\minute}
\def\draftlabel#1{{\@bsphack\if@filesw {\let\thepage\relax
   \xdef\@gtempa{\write\@auxout{\string
      \newlabel{#1}{{\@currentlabel}{\thepage}}}}}\@gtempa
   \if@nobreak \ifvmode\nobreak\fi\fi\fi\@esphack}
        \gdef\@eqnlabel{#1}}
\def\@eqnlabel{}
\def\@vacuum{}
\def\draftmarginnote#1{\marginpar{\raggedright\scriptsize\tt#1}}
\def\draft{\oddsidemargin -.2truein
        \def\@oddfoot{\sl preliminary draft \hfil
        \rm\thepage\hfil\sl\today\quad\militarytime}
        \let\@evenfoot\@oddfoot \overfullrule 3pt
        \let\label=\draftlabel
        \let\marginnote=\draftmarginnote
   \def\@eqnnum{(\theequation)\rlap{\kern\marginparsep\tt\@eqnlabel}%
\global\let\@eqnlabel\@vacuum}  }
\def\thebibliography#1{
\vskip 0.5cm \centerline{\bf \Large References}
\list{
[\arabic{enumi}]}{\settowidth\labelwidth{[#1]}
\leftmargin\labelwidth
\advance\leftmargin\labelsep
\usecounter{enumi}}
\def\newblock{\hskip .11em plus .33em minus .07em}
\sloppy\clubpenalty4000\widowpenalty4000
\sfcode`\.=1000\relax}

\renewcommand{\theequation}{\arabic{section}.\arabic{equation}}
\renewcommand{\section}{\setcounter{equation}{0}\@startsection
{section}{1}{0mm}{-\baselineskip}{0.5\baselineskip} {\normalfont\Large\bfseries}}
\renewcommand{\subsection}{\@startsection
{subsection}{2}{0mm}{-\baselineskip}{0.5\baselineskip} {\normalfont\large\bfseries}}
\renewcommand{\subsubsection}{\@startsection
{subsubsection}{3}{0mm}{-\baselineskip}{0.5\baselineskip}
{\normalfont\normalsize\slshape}}

\begin{document}


\begin{titlepage}
\begin{flushright}
CPHT-RR118.122018, December 2018
\vspace{1.5cm}
\end{flushright}
\begin{centering}
{\bm\bf \Large Spontaneous dark-matter mass generation along cosmological attractors in string theory}

\vspace{7mm}

 {\bf Thibaut Coudarchet\textsuperscript{1}, Lucien Heurtier\textsuperscript{2} \\and Herv\'e Partouche\textsuperscript{1}}

 \vspace{4mm}

{\textsuperscript{1}Centre de Physique Th\'eorique, Ecole Polytechnique,\footnote{Unit\'e  mixte du CNRS et de l'Ecole Polytechnique, UMR 7644.} \\ F--91128 Palaiseau, France\\ \textit{thibaut.coudarchet@polytechnique.edu, }\textit{herve.partouche@polytechnique.edu}}

\vspace{4mm}
{\textsuperscript{2}Department of Physics, University of Arizona, Tucson, AZ 85721\\ \textit{heurtier@email.arizona.edu}}

\end{centering}
\vspace{0.1cm}
$~$\\
\centerline{\bf\Large Abstract}\\
\vspace{-1cm}

\begin{quote}

\hspace{.6cm} 

We propose a new scenario for generating a relic density of non-relativistic dark matter in the context of heterotic string theory. Contrary to standard thermal freeze-out scenarios, dark-matter particles are abundantly produced while still relativistic, and then decouple from the thermal bath due to the sudden increase of their mass above the universe temperature. This mass variation is sourced by the condensation of an order-parameter modulus, which is triggered when the temperature $T(t)$ drops below the supersymmetry breaking scale $M(t)$, which are both time-dependent. A cosmological attractor mechanism forces this phase transition to take place, in an explicit class of heterotic string models with spontaneously broken supersymmetry, and at finite temperature.

\end{quote}

\end{titlepage}
\newpage
\setcounter{footnote}{0}
\renewcommand{\thefootnote}{\arabic{footnote}}
 \setlength{\baselineskip}{.7cm} \setlength{\parskip}{.2cm}

\setcounter{section}{0}


\section{Introduction}

As an ultraviolet complete theory unifying gravity with gauge interactions, string theory is a natural framework to study the primordial universe and describe cosmology using a top-down approach. In modern language, a possible question to be asked is whether  ingredients which are used in the most common cosmological scenarios --- such as models including a cosmological constant and a cold dark-matter component ($\Lambda$CDM) --- can  be derived from models of the string theory landscape, rather than   embedded in only apparently consistent low energy field theories of the swampland~\cite{landswamp}. 

It is nowadays established that our universe is constituted of three crucial components, which are dark energy, dark matter and Standard-Model particles. The amount  of each of these ingredients have been  measured with very good accuracy in the present universe~\cite{Aghanim:2018eyx}, indicating that a very large portion of the universe energy density is shared by dark energy and dark matter rather than baryons and radiation. Furthermore, the study of the cosmic microwave background  has shown to be compatible with dark energy and non-relativistic matter playing a key role in diluting the inhomogeneities of the primordial universe at early times, throughout a phase of so-called {\em cosmic inflation} (see \eg  Ref.~\cite{Baumann:2009ds} for a review).  If a lot  of the string-cosmology literature has been focusing on finding a way to generate enough $\mbox{$e$-folds}$ of inflation in the primordial universe (see debates on such a possibility~\cite{Obied:2018sgi,landswamp}), studies trying to obtain  a phase of matter domination during the late cosmological evolution of the universe are much more rare~\cite{Dashko:2018dsw}. 
In practice, most of the dark-matter models which have been proposed in the context of string theory are {\em string inspired}, in the sense that the particle interactions and mass spectrum are derived from string-theory models. Therefore, in such a framework,  the discussion of dark-matter decoupling and non-relativistic matter production remains to be an effective, low-energy discussion, or relies on purely geometrical effects such as domain-walls or cosmic-strings decay. The interesting possibility that a whole tower of KK-states contribute collectively to the dark-matter relic density, while different species decay at different time scales, was also proposed in Ref.~\cite{DDM} under the name {\em dynamical dark-matter}. In these models, the relic density is typically produced in the early universe through a misalignment mechanism. The freeze-out mechanism was also considered in Ref.~\cite{Dienes:2017zjq}, although in such context the particle spectrum is taken to be a time-independent data set, contrary to what we will consider.

In particular, in the usual {\em thermal freeze-out} scenario, it is assumed that a significant amount of dark matter is produced in the early universe, before it decouples from the thermal bath when the temperature drops under the dark-matter mass. In this paper, we present an alternative mechanism in which dark matter is naturally abundantly produced while still relativistic, and then decouples from the thermal bath due to the brutal variation of its mass above the temperature. This scenario arises  within a class of  explicit string models in $d$ dimensions, due to  cosmological attractors that yield a phase transition responsible for the spontaneous mass generation of the dark-matter particles. Note that the possibility of a variable-mass dark-matter particle has already  been proposed in Ref.~\cite{Franca:2004kk} in a different context, but  relatively unexplored from the phenomenological perspective.

In the past literature, heterotic string models compactified on tori (or orbifolds)  with spontaneously broken supersymmetry~\cite{SSstring,Kounnas-Rostand} \`a la Scherk-Schwarz~\cite{SS} have been considered at finite temperature~\cite{Kounnas-Rostand} and weak string coupling. It was shown that in the context of flat, homogenous and isotropic cosmological evolutions, the universe is attracted towards a ``radiation-like critical solution''~\cite{attractor, solcri, R4R5, cosmo_phases, cosmo_phases2}, along which the supersymmetry breaking scale $M(t)$, the temperature $T(t)$ and the inverse of the scale factor $a(t)$ evolve proportionally, $M(t)\propto T(t)\propto 1/a(t)$. The denomination ``radiation-like'' is motivated by the fact that the {\em total} energy density and pressure arising from $(i)$  the thermal bath of the infinite towers of Kaluza-Klein (KK) states along the internal Scherk-Schwarz directions  and ($ii$) the coherent motion of $M(t)$ satisfy the same state equation as pure radiation, $\rho_{\rm tot}=(d-1)P_{\rm tot}$~\cite{solcri,R4R5}. If helpful to understand the behavior of the early universe after reheating, when the light matter content of the universe is in thermal equilibrium, such a critical solution cannot be a low energy attractor for our universe since we know that ($i$) at present time the supersymmetry breaking scale is extremely large as compared to the universe temperature, and ($ii$) that the universe is matter dominated. Therefore, one needs to complexify the picture in order to open the possibility that part of the massless spectrum becomes massive 	and then decouples, while the universe evolves.

This is precisely what we do in the present work. We use the fact that at special points in moduli space, states  which are generically very heavy become massless~\cite{GV}. When these states contain more fermions than bosons, the free energy density $\F$ arising from their thermalized towers of KK modes shows very peculiar properties. First of all, at such a point in moduli space, $\F$ is extremal. Second, this extremum is a minimum (maximum) for large enough (low enough)  temperature $T$, as compared to the supersymmetry breaking scale~$M$. Assuming generic initial conditions compatible with a minimum, the destabilization of the order parameter, which is a modulus, then occurs dynamically, provided the attractor mechanism described in the above paragraph enforces  $T(t)/M(t)$ to reach low enough values. As a result, while the universe expands and the temperature (as well as the supersymmetry breaking scale) drops, for  the evolution dictated by the radiation-like critical solution to be approached, a phase transition takes place, where the condensation of the order-parameter modulus induces a large mass to the whole initially light KK towers. We will see that such a condensation is naturally pushed up to values which are necessarily larger than the temperature of the thermal bath, generating spontaneously an important amount of non-relativistic matter that may freeze-out later on \ie quit equilibrium, due to the expansion of the universe. 

The paper is organized as follows: In Sec.~\ref{S2},  we construct the simplest heterotic models for which the free energy density presents suitable features for developing the instability required for the spontaneous dark-matter mass generation. Sec.~\ref{transi1} is devoted to the analytical description of the attractor mechanisms. In a first stage, the order-parameter modulus is attracted towards the minimum of its potential well, while the whole cosmological evolution approaches a radiation-like critical solution~\cite{cosmo_phases,cosmo_phases2}. This effect is already non-trivial, in the sense that the mechanism avoids the so-called ``cosmological moduli problem''~\cite{cosmomodprob}.\footnote{Typically encountered in inflationary scenarios,  the universe at intermediate times may be dominated by the energy stored in massive scalars, which cannot stabilize. Their eventual decay into radiation can lead to an entropy excess.} Then, the ratio $T/M$ being dynamically pushed to some low enough value, the phase transition suddenly takes place, inducing the dark-matter particle to become heavier than the temperature scale. In Sec.~\ref{dark}, we first review how dark-matter particles decouple from a thermal bath in the context of the usual thermal freeze-out scenario. Then, we present the new mechanism which we propose in this paper to make dark matter decouple spontaneously from the thermal bath, when the phase transition happens. We finally relate the relic energy density of cold dark matter to the scale factor of the universe and the freshly acquired dark-matter mass.   Our conclusions and perspectives can be found in Sec.~\ref{conclu}, where we summarize our results and present futur prospects.  


\section{Thermal effective potential}
\label{S2}

Throughout this paper, all  dimensionful quantities will be expressed in string units $\mbox{$(\alpha'=1)$}$, and denoted with suffixes ``$(\sigma)$'' when measured in string frame \ie $\sigma$-model frame.
In this section, we consider models realizing a spontaneous breaking of supersymmetry at a scale $M_{(\sigma)}$, and derive their free energy density $\F_{(\sigma)}$ at finite temperature $T_{(\sigma)}$.  
To be more specific, we would like $\F_{(\sigma)}$, which is nothing but the effective potential at finite temperature, to depend on a modulus that will be massive at high temperature and tachyonic at low temperature, as compared to the supersymmetry breaking scale. As will be shown in Sec.~\ref{transi1}, the dynamics of the universe may then enforce the time evolutions of $T_{(\sigma)}$ and $M_{(\sigma)}$ to trigger a destabilization of the modulus, which is responsible for a dark-matter mass generation.  


\subsection{Heterotic models and free energy}

Our starting point is the $E_8\times E_8$ heterotic string compactified on the  background
\be
S_{\text{E}}^1(R_{0})\times \R^{d-1}\times T^2\times T^{8-d},
\ee
where time is Euclidean and compactified on a circle of radius $R_0$, and $\R^{d-1}$ stands for the spatial directions. For simplicity, we consider the internal space to be factorized in two tori. The radius of one direction in $T^2$, say $X^d$, is the modulus to be (de-)stabilized, while the second direction, which we denote by $X^9$, is responsible for the spontaneous breaking of supersymmetry. On the contrary, all moduli associated with $T^{8-d}$ will play a minor role in the sequel. 

Technically, both finite temperature and spontaneous breaking of supersymmetry can be implemented by a stringy version of the Scherk-Schwarz mechanism~\cite{SSstring,Kounnas-Rostand}. At 1-loop, the free energy density can be written as
\be
\begin{aligned}
\F_{(\sigma)}= -{1\over 2(2\pi)^d}&\int_F {\dd\tau_1\dd\tau_2\over \tau_2^{1+{d+2\over 2}}}\, \sum_{g_0,h_0}\sum_{\tilde k_0,l_0}e^{-{\pi R_0^2\over \tau_2}|2\tilde k_0+g_0+(2l_0+h_0)\tau|^2}\\ 
&{1\over 2}\sum_{g_9,h_9}{\sqrt{\det G}}\sum_{\overset{\scriptstyle \tilde k_d, l_d}{\tilde k_9, l_9}}e^{-{\pi\over \tau_2}\left[\tilde k_i+{g_i\over 2}+(l_i+{h_i\over 2})\bar \tau\right](G_{ij}+B_{ij})\left[\tilde k_j+{g_j\over 2}+(l_j+{h_j\over 2})\tau\right]}\\
& {1\over 2}\sum_{a,b}(-1)^{a+b+ab}\theta[^a_b]^4\, (-1)^{g_0a+h_0b+g_0h_0}\, (-1)^{g_9a+h_9b+g_9h_9}\\
& {1\over 2}\sum_{\gamma,\delta}\bar \theta[^\gamma_\delta]^8\, (-1)^{g_9\gamma+h_9\delta+g_9h_9}\; {1\over 2}\sum_{\gamma',\delta'}\bar \theta[^{\gamma'}_{\delta'}]^8\, (-1)^{g_9\gamma'+h_9\delta'+g_9h_9}\\
&{\Gamma_{8-d,8-d}\over \eta^{12}\bar \eta^{24}},
\end{aligned}
\label{free1}
\ee
where we use the following notations:

$\bullet$  $\tau=\tau_1+i\tau_2$ is the Teichm\"uller parameter of the genus-1 Riemann surface and $F$ the fundamental domain of the modular group. $\eta(\tau)$ and $\theta [^\alpha_\beta](\tau)$ are the Dedekind and Jacobi modular forms, for which conventions can be found in Ref.~\cite{KiritsisBook}.

$\bullet$ The lattices of zero modes associated to the Euclidean circle and the $T^2$ coordinates are in the first and second lines. The numbers $\tilde k_0,\tilde k_d,\tilde k_9$ and $l_0,l_d,l_9$ are arbitrary integers, while $g_0, h_0$ and $g_9,h_9$ are parities \ie equal to 0 or 1. For notational compactness, we have also introduced $g_d,h_d$ but those are simply vanishing. Moreover, $G_{ij}$ and $B_{ij}$ are the metric and antisymmetric tensor background values on $T^2$, to be specified shortly.  

$\bullet$ The worldsheet left-moving fermions contribute to the conformal block in the third line. The latter is dressed with ``cocycles'' \ie phases that couple the above mentioned lattices to the spin structures $a,b\in\{0,1\}$, thus implementing finite temperature and spontaneous breaking of supersymmetry~\cite{Kounnas-Rostand}. In string frame, the temperature is the inverse of the Euclidean-time circle circumference, 
\be
T_{(\sigma)}={1\over 2\pi R_0}.
\ee

$\bullet$ In the fourth line, the 16 extra right-moving coordinates of the bosonic string yield two $E_8$ lattices, where $\gamma,\delta$ and $\gamma',\delta'\in\{0,1\}$. Cocycles responsible for the $E_8\times E_8\to SO(16)\times SO(16)$ spontaneous breaking  are also included~\cite{o16}. In total, the lattice of the direction $X^9$ is involved in the phase
\be
(-1)^{g_9(a+\gamma+\gamma')+h_9(b+\delta+\delta')+g_9h_9},
\ee
which shows that super-Higgs and Higgs mechanisms combine in a non-trivial way. Consider an initially massless, supersymmetric pair of bosonic ($a=0$) and fermionic ($a=1$) degrees of freedom: If their gauge charge $\gamma+\gamma'$ is even, then the Scherk-Schwarz mechanism along $X^9$ induces a non-trivial mass to the fermion, while the boson remains massless.  On the contrary, when $\gamma+\gamma'$ is odd, the mass splitting is reversed, in the sense that the boson becomes massive, while the fermion remains massless~\cite{solcri, R4R5, Kounnas:2016gmz}.  

$\bullet$ The last line contains the lattice of zero modes associated to the internal directions $X^{d+1},\dots, X^8$, and worldsheet left- or right-moving oscillator contributions.  

$\bullet$ We consider a $T^2$ metric and antisymmetric tensor 
\be
\label{background}
(G+B)_{ij}=\begin{pmatrix}R_d^2 & \epsilon \\-\epsilon & 4R_9^2\end{pmatrix}\!, \quad\;\; i,j\in\{d,9\},
\ee
where $R_d$ and $R_9$ are dynamical radii, while $\epsilon\in\Z$ is a constant background. To motivate this choice, notice that in the absence of any cocycle responsible for finite temperature and supersymmetry breaking along $X^9$, we would have an $U(1)\to SU(2)$  enhancement of the gauge symmetry at $R_d=1$ and arbitrary $\epsilon\in\Z$.  In fact, a pair of non-Cartan vector multiplets would be exactly massless at such a point in moduli space. As shown in great details in Ref.~\cite{CP}\footnote{In the appendix of Ref.~\cite{CP}, all marginal deformations of the heterotic theory are taken into account. However, for the sake of clarity and simplicity in the present work, we only discuss and keep dynamical the moduli relevant to the phase transition under consideration.}, once  supersymmetry breaking is implemented along $X^9$, the effect of an even value of the ``discrete Wilson line'' $\epsilon$ is to induce a tree-level mass  $1/(2R_9)$ (equal to that of the gravitini) only to the fermions of the non-Cartan vector multiplets. Conversely,  an odd value of $\epsilon$ implies the fermions to remain massless, while their bosonic superpartners  become massive.  In both cases, we may define the scale of supersymmetry breaking in string frame to be
\be
M_{(\sigma)}={1\over 2\pi R_9}.
\ee
In the remaining part of this subsection, we show how the picture is generalized in presence of both supersymmetry breaking and finite temperature.  

Redefining $a=\hat a+h_0+h_9$, $b=\hat b+g_0+g_9$, and using the Jacobi identity to handle the sum over $\hat a, \hat b$,  the free energy density becomes
\be
\begin{aligned}
\F_{(\sigma)}=&\,  {1\over 2(2\pi)^d}\int_F {\dd\tau_1\dd\tau_2\over \tau_2^{1+{d+2\over 2}}}\, \sum_{g_0,h_0}\sum_{\tilde k_0,l_0}e^{-{\pi R_0^2\over \tau_2}|2\tilde k_0+g_0+(2l_0+h_0)\tau|^2}\\ 
&\, R_d\sum_{\tilde m_d, n_d}e^{-{\pi R_d^2\over \tau_2}|\tilde m_d+n_d\tau|^2}\, R_9\sum_{g_9,h_9}\sum_{\tilde k_9, l_9}e^{-{\pi R_9^2\over \tau_2}|2\tilde k_9+g_9+(2l_9+h_9)\tau|^2}\, (-1)^{\epsilon(\tilde m_dh_9+n_dg_9)}\\
&\,  \theta\big[{}^{1-h_0-h_9}_{1-g_0-g_9}\big]^4\, {1\over 2}\sum_{\gamma,\delta}\bar \theta[^\gamma_\delta]^8\; {1\over 2}\sum_{\gamma',\delta'}\bar \theta[^{\gamma'}_{\delta'}]^8\, {\Gamma_{8-d,8-d}\over \eta^{12}\bar \eta^{24}}\, (-1)^\varphi,\\
\where\quad &\, \varphi = g_{0}+g_{9}+h_{0}+h_{9}+g_{9}h_{0}+g_{0}h_{9}+g_{9}(\gamma+\gamma')+h_{9}(\delta+\delta').
\end{aligned}
\label{free2}
\ee
To proceed, we assume that the radii of the periodic directions $X^0$, $X^9$, are large compared to the Hagedorn radius, in order for Hagedorn-like instabilities not to occur,
\be
R_0, \; R_9\gg R_H={1+\sqrt{2}\over \sqrt{2}}.
\ee 
This guarantees that the integrand does not develop level-matched tachyonic modes and the free energy to be well defined. By noticing that all contributions with non-vanishing winding numbers $2l_0+h_0$ or $2l_9+h_9$ yield contributions $\O(e^{-\# R_0^2})$ or $\O(e^{-\# R_9^2})$, where $\#$ is positive and $\O(1)$, we may focus on the sectors $h_0=h_9=0$, with $l_0=l_9=0$. Due to the $\theta\big[{}^{\ \ \ \ 1}_{1-g_0-g_9}\big]^4$ factor, non-trivial contributions arise only for $(g_0,g_9)=(1,0)$ or $(0,1)$. As a result, we obtain
\be
\begin{aligned}
\!\!\F_{(\sigma)}\!=&\,  {R_9\over 2(2\pi)^d}\int_F {\dd\tau_1\dd\tau_2\over \tau_2^{1+{d+1\over 2}}}\!\!\! \sum_{\overset{\scriptstyle (g_0,g_9)=}{(1,0)\, {\rm or}\, (0,1)}}\sum_{\tilde k_0, \tilde k_9}e^{-{\pi \over \tau_2}\left[R_0^2(2\tilde k_0+1)^2+R_9^2(2\tilde k_9+1)^2\right]}\, \sum_{m_d,n_d}q^{{1\over 2}p_L^2}\bar q^{{1\over 2}p_R^2}\,  (-1)^{\epsilon n_dg_9}\\
&\,  {\theta\big[{}^1_0\big]^4\over \eta^{12}\bar \eta^{24}}\, {1\over 2}\sum_{\gamma,\delta}\bar \theta[^\gamma_\delta]^8\, (-1)^{g_9\gamma}\; {1\over 2}\sum_{\gamma',\delta'}\bar \theta[^{\gamma'}_{\delta'}]^8\, (-1)^{g_9\gamma'}\, \Gamma_{8-d,8-d}+\O(e^{-\# R_0^2}) + \O(e^{-\# R_9^2}),
\end{aligned}
\label{free3}
\ee
where $q=e^{2i\pi \tau}$. In this expression, we have written the lattice of zero modes associated to $S^1(R_d)$ in Hamiltonian form, where 
\be
p_L={1\over \sqrt{2}}\Big({m_d\over R_d}+n_d R_d\Big)^2, \quad p_R={1\over \sqrt{2}}\Big({m_d\over R_d}-n_d R_d\Big)^2.
\ee
Due to the presence of factors $e^{-{\pi R_0^2\over \tau_2}(2\tilde k_0+1)^2}$ or $e^{-{\pi R_9^2\over \tau_2}(2\tilde k_9+1)^2}$ in the integrand, we may extend the fundamental domain $F$ of integration to the ``upper half strip'',
\be
\int_F{\dd\tau_1\dd\tau_2}\, (\, \cdots)= \int_{-{1\over 2}}^{1\over 2}\dd\tau_1\int_0^{+\infty}\dd\tau_2\, (\, \cdots)+\O(e^{-\# R_0^2}) + \O(e^{-\# R_9^2}).
\ee
Hence, integrating over $\tau_1$ projects on the physical \ie level-matched spectrum. 

To evaluate explicitly the free energy, we expand
\be
\label{expand}
\begin{aligned}
&{\theta_2^4\over \eta^{12}}=16\big(1+\O(q)\big), \\
 &{1\over 2}\sum_{\gamma,\delta}\bar \theta\big[{}^\gamma_\delta\big]^8(-1)^{g_9\gamma}=1+112\bar q+(-1)^{g_9} 128 \bar q+\O(\bar q^2),\\
& {1\over \bar \eta^{24}}={1\over \bar q}\big(1+24\bar q+\O(\bar q^2)\big).
\end{aligned}
\ee
Moreover, choosing the radius of the direction $X^d$ to be ``moderate'', 
\be
{1\over R_0}, {1\over R_9}\ll R_d\ll R_0, R_9, 
\label{Rrange}
\ee
$R_d$ may sit in the neighborhood of 1, where the states with momenta and winding numbers $m_d=-n_d=\pm 1$ become massless. We may then write
\be
\sum_{m_d,n_d}q^{{1\over 2}p_L^2}\bar q^{{1\over 2}p_R^2}\,  (-1)^{\epsilon n_dg_9} = 1+2(-1)^{\epsilon g_9}\bar q\, e^{-\pi\tau_2\left(R_d-{1\over R_d}\right)^2}+\cdots, 
\ee
where the ellipses stand for all other modes,  $m_dn_d\neq -1$. Note that the latter cannot yield states in the spectrum simultaneously level-matched and lighter than $T_{(\sigma)}$ and $M_{(\sigma)}$. 
In a similar way, we  assume the size of $T^{8-d,8-d}$ to be  ``moderate'', \ie with metric satisfying
\be
{1\over R_0^2}, {1\over R_9^2}\ll|G_{IJ}|\ll R_0^2, R_9^2, \qquad I,J\in\{d+1,\dots,8\}.
\label{range}
\ee
Hence, $(G+B)_{IJ}$ may sit at a point of enhanced gauge symmetry in moduli space, $U(1)^{8-d}\to G_{\rm en}$, so that 
\be
 \Gamma_{(8-d,8-d)}=1+d_{\rm en}\bar q +\cdots.
 \ee
In the above formula, we take for simplicity $(G+B)_{IJ}$ to sit exactly at such a point, or to be outside of their neighborhoods, in which case   $d_{\rm en}=0$. 
We are now ready to integrate physical mode by physical mode. This can be done using the identity
\be 
\label{dexpsup}
\Hc_\nu(x)\equiv {1\over \Gamma(\nu)}\int_0^{+\infty}{{\dd u}\over u^{1+\nu}}\, e^{-{1\over u}-x^2u}={2\over \Gamma(\nu)}\, x^\nu K_\nu(2x),
\ee
where $K_\nu$ is the modified Bessel function of the second kind.  In practice, $x$ is essentially the ratio of mass in the spectrum, to $T_{(\sigma)}$ or $M_{(\sigma)}$. As a consequence, different contributions can have different orders of magnitude, as follows from the behaviors of $\Hc_\nu(x)$ at large and small arguments, 
\be 
\label{expsup}
\begin{aligned}
&\Hc_\nu(x)\sim {\sqrt{\pi}\over \Gamma(\nu)}\, x^{\nu-{1\over 2}}\, e^{-2x} \,\;\;\;\;\;\quad\mbox{when} \quad x\gg 1,\\
&\Hc_\nu(x)=1-{x^2\over \nu-1}+\O(x^4)\quad\mbox{when}\quad |x|\ll 1.
\end{aligned}
\ee
The dominant contribution to $\F_{(\sigma)}$ arises from the (nearly) massless states, including those with $m_d=-n_d=\pm1$ when $R_d\simeq 1$, together with their  towers of KK modes associated to the Euclidean time and direction $X^9$. All other states yield exponentially suppressed contributions. They include in particular those arising from oscillator modes at the string scale, or  from the states winding around the large compact directions $X^0$, $X^9$. 

To write the final result, it is convenient to define 
\be
\zeta=\ln(R_{d}),
\qquad
\eta=\ln(R_{9}),
\qquad
z=\ln\!\left(\frac{R_{0}}{R_{9}}\right)\!=\ln\!\left(\frac{M_{(\sigma)}}{T_{(\sigma)}}\right)\!,
\ee
in terms of which we find
\be
\F_{(\sigma)}=T_{(\sigma)}^d f(z,\eta,\zeta)+\O\!\left((c\Ms T_{(\sigma)})^{d\over 2}e^{-c\Ms/T_{(\sigma)}}\right)+\O\!\left((c\Ms M_{(\sigma)})^{d\over 2}e^{-c\Ms/M_{(\sigma)}}\right)\!.
\label{Ftot}
\ee
In this expression, $c\Ms>0$ is the lowest (Higgs-like) mass scale generated by the moduli $G_{IJ}$. As follows from Eq.~(\ref{range}),  it is heavier than $T_{(\sigma)}$ and $M_{(\sigma)}$, thus yielding exponential suppression.\footnote{For instance, if $G_{IJ}=\O(1)$, then so is $c$.} The dominant contribution in $\F_{(\sigma)}$ involves 
\be
\begin{aligned}
f(z,\eta,\zeta)=&-(\nF+\nB)f^{(d)}_{\rm T}(z)+(\nF-\nB)f^{(d)}_{\rm V}(z)\\
&-( \nFt+ \nBt)\tilde f^{(d)}_{\rm T}(z,\eta,\zeta)+( \nFt- \nBt)\tilde f^{(d)}_{\rm V}(z,\eta,\zeta),
\end{aligned}
\label{f}
\ee
where $\nB$ and $\nF$ are the numbers of  bosonic and fermionic massless states for generic $R_d$, while $\nBt$ and $\nFt$ count those becoming massless at $R_d=1$. The dressing functions account for the corresponding towers of KK modes along $X^0$, $X^9$~\cite{cosmo_phases,cosmo_phases2},
\be
\begin{aligned}
&f^{(d)}_{\text{T}}(z)=\frac{\Gamma\!\left(\frac{d+1}{2}\right)}{\pi^{\frac{d+1}{2}}}\sum_{\tilde k_{0},\tilde k_{9}}\frac{e^{dz}}{\left[e^{2z}(2\tilde k_{0}+1)^2+(2\tilde k_{9})^2\right]^{\frac{d+1}{2}}},\\
&\tilde f^{(d)}_{\text{T}}(z,\eta,\zeta)=\frac{\Gamma\!\left(\frac{d+1}{2}\right)}{\pi^{\frac{d+1}{2}}}\sum_{\tilde k_{0},\tilde k_{9}}\frac{e^{dz}\, \Hc_{\frac{d+1}{2}}\Big(\pi |e^\zeta-e^{-\zeta}|e^\eta\sqrt{e^{2z}(2\tilde k_{0}+1)^2+(2\tilde k_{9})^2}\Big)}{\left[e^{2z}(2\tilde k_{0}+1)^2+(2\tilde k_{9})^2\right]^{\frac{d+1}{2}}},
\end{aligned}
\ee
while the last two functions can be defined by 
\be
f^{(d)}_{\text{V}}(z)\equiv e^{(d-1)z}f^{(d)}_{\text{T}}(-z), \qquad \tilde f^{(d)}_{\text{V}}(z,\eta,\zeta)\equiv e^{(d-1)z}\tilde f^{(d)}_{\text{T}}(-z,\eta+z,\zeta).
\ee

To be specific, the massless spectrum satisfies
\be
\begin{aligned}
&\nB = 8\times (8+120+120+d_{\rm en}),&&\nF=8\times(128+128),\\
&\nBt = 8\times 2\, (1-\epsilon),&&\nFt=8\times 2\,\epsilon.
\end{aligned}
\ee
There is a universal degeneracy factor 8 arising from the fact that at zero temperature and supersymmetry breaking scale, the theory is maximally supersymmetric \ie with 16 supercharges ($\N=4$ in 4 dimensions). In $\nB$, the $8\times 8$ degrees of freedom are those of the metric, antisymmetric tensor and dilaton field dimensionally reduced from  10 to $d$ dimensions. The 120's are the dimensions of the two $SO(16)$ gauge groups, while $d_{\rm en}$ is the number of roots of the enhanced $U(1)^{8-d}\to G_{\rm en}$ gauge factor. They all satisfy $(\gamma,\gamma')=(0,0)$. In $\nF$, the 128's are the dimensions of the spinorial representations of the $SO(16)$ factors, corresponding to $(\gamma,\gamma')=(1,0)$ and $(0,1)$. At $R_d=1$, modes having $m_d=-n_d=\pm1$ are massless, with charges $p_R=\pm \sqrt{2}$ under the right-moving $U(1)$ isometry group of $S^1(R_d)$. Either $\nBt$ or $\nFt$ is non-trivial: When $\epsilon=0$, the modes are bosons corresponding to the roots of the enhanced $U(1)\to SU(2)$ and on the contrary, $\epsilon=1$ yields a pair of fermionic  states, charged under $U(1)$ which is not enhanced. 


\subsection{Properties of the free energy}

From now on, we will neglect in $\F_{(\sigma)}$ (and omit in all formulas) the exponentially-suppressed contributions in Eq.~(\ref{Ftot}).  Some remarks are in order:

$\bullet$ $\F_{(\sigma)}$ is the free energy density valid for arbitrary mass $|R_d-1/R_d|$ of the $\nBt$ or $\nFt$ states, provided Eq.~(\ref{Rrange}) holds. Consistently, we find that when they are massless, \ie at $R_d=1$,  
\be
\begin{aligned}
&f(z,\eta,0)=-(N_{\rm F}+N_{\rm B})f^{(d)}_{\rm T}(z)+(N_{\rm F}-N_{\rm B})f^{(d)}_{\rm V}(z),\\
\where \quad &N_{\rm F}=\nF+\nFt,\quad N_{\rm B}=\nB+\nBt.
\end{aligned}
\ee

$\bullet$ The above split of $f$ into two pieces is motivated by taking the  limit $z\to +\infty$, where thermal effects are screened by quantum effects. In fact, we have 
\be
\F_{(\sigma)}|_{R_d=1} \underset{z\to+\infty}\sim M_{(\sigma)}^d\,  (N_{\rm F}-N_{\rm B})\, \xi,\quad \where \quad \xi={\Gamma\big({d+1\over 2}\big)\over \pi^{d+1\over 2}}\,  \sum_m{1\over | 2m+1|^{d+1}},
\label{a1}
\ee
which reproduces the expression of the 1-loop effective potential at zero temperature in a theory where supersymmetry is spontaneously broken by the Scherk-Schwarz mechanism~\cite{Abel:2015oxa,solcri,R4R5}. 

$\bullet$ Conversely, when $z\to -\infty$, quantum corrections are screened by thermal effects. As a result, we  recover
\be
\F_{(\sigma)}|_{R_d=1} \underset{z\to-\infty}\sim-\, 2\pi R_9\, T_{(\sigma)}^{d+1}\, (N_{\rm F}+N_{\rm B})\,  \xi,
\label{a2}
\ee
which is the Stefan-Boltzmann law for radiation in $d+1$ dimensions. The overall factor $2\pi R_9$ arises consistently with the interpretation of the density in $d+1$ dimensions. 

\begin{figure}[!h]
\vspace{.4cm}
\begin{center}
\includegraphics[trim=0cm 6.5cm 0cm 0cm,clip,scale=0.65]{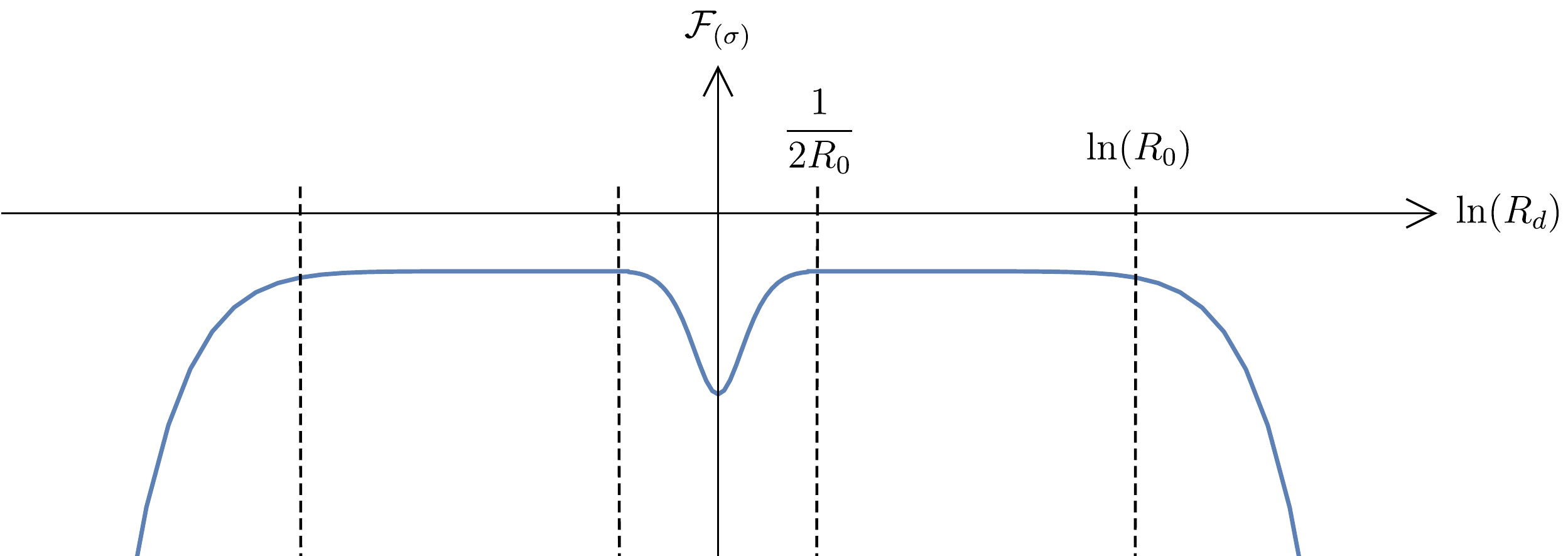}\end{center}
\caption{\footnotesize \em Qualitative shape of the the free energy density $\F_{(\sigma)}$ as a function of $\zeta=\ln R_d$, when $\nFt-\nBt\le 0$, or for low enough $M_{(\sigma)}/T_{(\sigma)}$ when $\nFt-\nBt> 0$. Several phases can be identified: A well of size $\max (1/R_0,1/R_9)$. On both of its sides, plateaus extend until $\pm\min(\ln R_0,\ln R_9)$. 
The latter are followed/preceded by exponential falls if $\nF-\nB \le 0$, 
or for low enough $M_{(\sigma)}/T_{(\sigma)}$ when $\nF-\nB> 0$. The exponential behavior is increasing for large enough $M_{(\sigma)}/T_{(\sigma)}$, when $\nF-\nB> 0$.}
\label{well}
\end{figure}

\begin{figure}[!h]
\vspace{.4cm}
\begin{center}
\includegraphics[trim=0cm 6.5cm 0cm 0cm,clip,scale=0.65]{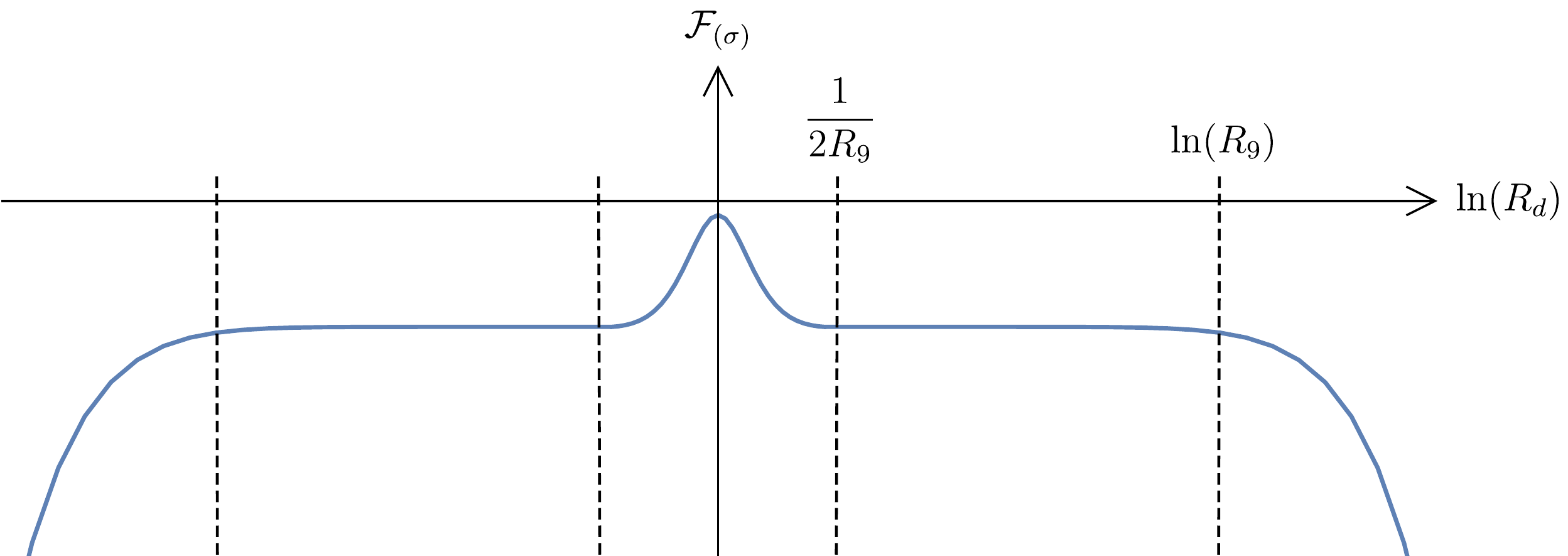}\end{center}
\caption{\footnotesize \em Qualitative shape of the the free energy density $\F_{(\sigma)}$ as a function of $\zeta=\ln R_d$, in the case $\nFt-\nBt> 0$,  when $M_{(\sigma)}/T_{(\sigma)}$ is large enough. Several phases can be identified: A bump of size $\max (1/R_0,1/R_9)$. On both of its sides, plateaus extend until $\pm\min(\ln R_0,\ln R_9)$. 
The latter are followed/preceded by exponential falls if $\nF-\nB \le 0$, 
or for low enough $M_{(\sigma)}/T_{(\sigma)}$ when $\nF-\nB> 0$. The exponential behavior is increasing for large enough $M_{(\sigma)}/T_{(\sigma)}$, when $\nF-\nB> 0$.}
\label{bump}
\end{figure}

$\bullet$ Expanding around $\zeta=0$, we identify the mass term of $\zeta$, 
\be
\begin{aligned}
f(z,\eta,\zeta)=&-(N_{\rm F}+N_{\rm B})f^{(d)}_{\rm T}(z)+(N_{\rm F}-N_{\rm B})f^{(d)}_{\rm V}(z)\\
&+{\zeta^2\over \pi T_{(\sigma)}^2}\!\left[(\nFt+\nBt)f^{(d-2)}_{\rm T}(z)-(\nFt-\nBt)f^{(d-2)}_{\rm V}(z)\right]\!+\O(\zeta^4),\\
\where \;\;\;\;\quad &\!\!\!\nFt+\nBt=8\times 2, \quad -(\nFt-\nBt)=(-1)^\epsilon\, 8\times 2.
\end{aligned}
\label{mas}
\ee 
When the extra massless states at $R_d=1$  are bosons ($\epsilon$  even), $\zeta$ is massive. Thus,  as a function of $\zeta$, $\F_{(\sigma)}$ presents a local minimum at $\zeta=0$, as shown in the qualitative Fig.~\ref{well}. 
However,  the situation is more involved  when the  massless states are fermions ($\epsilon$ odd). By noticing that the limits $z\to +\infty$ and $-\infty$ of the functions $f_{\rm T}^{(d)}(z)$ and   $f_{\rm V}^{(d)}(z)$ that we took in Eqs~(\ref{a1}) and~(\ref{a2}) can be easily applied to the $d-2$ case, we conclude that for large enough $e^z$, $\zeta$ is tachyonic, while for low enough $e^z$, it is massive. The tachyonic case is illustrated in Fig.~\ref{bump}, 
where $\F_{(\sigma)}$ has a maximum at $\zeta=0$. The massive case is as before, shown in Fig.~\ref{well}.  
The dynamical switch from the massive case to the tachyonic case will be used in the next section to trigger the destabilization of $R_d$, which is responsible for the mass generation of fermionic dark matter. 

$\bullet$ For $R_d$ sufficiently far from the self-dual point, the masses of the $\nBt$ or $\nFt$ states (depending on the parity of $\epsilon$) exceed $T_{(\sigma)}$ and $M_{(\sigma)}$. Thus, their contributions to the free energy become  exponentially suppressed and the second line of Eq.~(\ref{f}), which captures all $\zeta$- and $\eta$-dependences, can be omitted. Hence, as a function of $\zeta$, $\F_{(\sigma)}$ develops a well or a bump around $\zeta=0$, whose size is $\max(1/R_0,1/R_9)$, and on both sides of which is a plateau (see Figs~\ref{well},~\ref{bump}). 


\paragraph{Large extra dimension regime:}

For completeness, we may ask what is the behavior of the free energy  when the condition~(\ref{Rrange}) is relaxed. When $R_d\gtrsim R_0$ or $R_9$, KK modes along $X^d$ are lighter than $T_{(\sigma)}$ or   $M_{(\sigma)}$ and their contributions to the free energy are no more exponentially suppressed. Similarly, the winding modes along $S^1(R_d)$ start contributing to the free energy when $R_d\lesssim 1/R_0$ or $1/R_9$. On the contrary, the $\nBt$ and $\nFt$ states being even heavier than when $\zeta$ sits on a plateau discussed above, they can be omitted in the evaluation of $\F_{(\sigma)}$. Under such conditions, one obtains~\cite{cosmo_phases}
\begin{align}
f(z,\eta,\zeta)&= -(\nF+\nB)\big[f^{(d)}_{\rm T}(z)+k^{(d)}_{\rm T}(z,\eta-|\zeta|)\big]+(\nF-\nB)[f^{(d)}_{\rm V}(z)+k^{(d)}_{\rm V}(z,\eta-|\zeta|)\big]\espD\nonumber \\
& = e^{|\zeta|-\eta-z}\left[-(\nF+\nB) F^{(d+1)}_{\rm T}(z,\eta-|\zeta|)+(\nF-\nB)F^{(d+1)}_{\rm V}(z,\eta-|\zeta|)\right]\!,
\label{f2}
\end{align}
where, in the first line, the functions $k_{\rm T}^{(d)}$ and $k_{\rm V}^{(d)}$ account for the additional corrections attributed to the KK or winding states,  
\begin{equation}
\begin{aligned}
k_{\rm T}^{(d)}(z,\eta-|\zeta|)&=\frac{\Gamma\!\left(\frac{d+1}{2}\right)}{\pi^{\frac{d+1}{2}}}\sum_{m_{d}\neq 0}\sum_{\tilde{k}_{0},\tilde{k}_{9}}\frac{e^{dz}\, \mathcal{H}_{\frac{d+1}{2}}\Big(\pi |m_{d}|e^{\eta-|\zeta|}\sqrt{e^{2z}(2\tilde{k}_{0}+1)^{2}+(2\tilde{k}_{9})^{2}}\Big)}{\left[e^{2z}(2\tilde{k}_{0}+1)^{2}+(2\tilde{k}_{9})^{2}\right]^{\frac{d+1}{2}}},\\ 
k^{(d)}_{\rm V}(z,\eta-|\zeta|) &=e^{(d-1)z}~k^{(d)}_{\rm T}(-z,\eta-|\zeta|+z) .
\end{aligned}
\end{equation}
In the second line of Eq.~(\ref{f2}), a Poisson summation on the momentum  (or winding number) along $S^1(R_d)$ is performed, which yields 
\be
\label{fkgjT}
\begin{aligned}
F^{(d+1)}_{\rm T}(z,\eta-|\zeta|)&=\displaystyle{\Gamma\!\left({d+2\over 2}\right)\over \pi^{d+2\over 2}}\sum_{\tilde k_0,\tilde k_9,\tilde m_d}{e^{(d+1)z}\over \left[e^{2z}(2\tilde k_0+1)^2+(2\tilde k_9)^2+e^{-2(\eta-|\zeta|)}\tilde m_d^2\right]^{d+2\over 2}},\\
F^{(d+1)}_{\rm V}(z,\eta-|\zeta|)&=e^{dz}F^{(d+1)}_{\rm T}(-z,\eta-|\zeta|+z).
\end{aligned}
\ee
Much of the behavior of the free energy is captured in the regime $R_d\gg R_0, R_9$ (or $R_d\ll  1/R_0, 1/R_9$), which can be derived from the second expression in Eq.~(\ref{f2}). Defining $u=1$ or $-1$ to treat both cases simultaneously, we obtain  
\be\label{eq:FLargeExtraDim}
\F_{(\sigma)}=2\pi R_d^u\, T_{(\sigma)}^{d+1}\!\left[-(\nF+\nB)f^{(d+1)}_{\rm T}(z)+(\nF-\nB)f^{(d+1)}_{\rm V}(z)-\Big({R_0\over  R_d^u}\Big)^dn_{\rm B}\xi'+\cdots\right]\!,
\ee
where the ellipses stand for exponentially suppressed terms in $R_d^u/R_0$ and $R_d^u/R_9$, and 
\be
\xi'={\Gamma\big({d\over 2}\big)\over \pi^{d\over 2}}\,  \zeta(d).
\ee
The factor $2\pi R_d^u$ of Eq.~\eqref{eq:FLargeExtraDim} may be used to interpret the result in $d+1$ dimensions. However,  from the $d$-dimensional point of view, this translates into an exponential behavior, $\F_{(\sigma)}\propto e^{|\zeta|}$ for $|\zeta|\to \infty$, which we took to be decreasing  in Figs~\ref{well} and \ref{bump}. If this is so when $\nF-\nB\le 0$, this is not always true when $\nF-\nB>0$. In the latter case, we can use for $d+1$ the limits $z\to \pm \infty$ taken in Eqs~(\ref{a1}) and~(\ref{a2}) to conclude that the exponential behavior is decreasing for low enough $e^z$, and increasing  for large enough $e^z$. 


\section{Dynamical stabilization $\boldsymbol{/}$ destabilization}
\label{transi1}

In this section, we consider the free energy $\F_{(\sigma)}=T_{(\sigma)}^df$ in generic models, namely with the function $f$ given in Eq.~(\ref{f}) (or~(\ref{f2})), for arbitrary $\nF$, $\nB$ and $\nFt$, $\nBt$. We want to show how the nature of the $\nFt+\nBt$ states becoming massless at $R_d=1$ impacts the dynamics and final expectation value of $R_d$. After deriving the cosmological equations of motion, we review the evolution found for $\nFt-\nBt < 0$, which  was considered in Ref.~\cite{cosmo_phases} and yields a stabilization of the modulus at the self-dual point. Then, we turn to our main case of interest, namely  $\nFt-\nBt>0$, which can trigger dynamically the destabilization of $R_{d}$ from its self-dual point. During this process,  the $\nFt+\nBt$ initially massless states acquire a large mass.  Becoming non-relativistic, we will see in Sect.~\ref{dark} that they may realize a component of cold dark matter in our universe, given that they are stable on cosmological time scales.


\subsection{Equations of motion and thermodynamics}
\label{seq}

Our starting point is the 1-loop effective action in $d$ dimensions. Considering only the degrees of freedom relevant for the (de-)stabilization mechanism, we have 
\be
\S=\int\dd^{d}x\sqrt{-g_{(\sigma)}}\left[e^{-2\phi}\!\left(\frac{\cR_{(\sigma)}}{2}+2\, \partial_{\mu}\phi\partial^{\mu}\phi-\frac{1}{2}{\partial_{\mu}R_9\partial^{\mu}R_9\over R_9^2}-\frac{1}{2}{\partial_{\mu}R_d\partial^{\mu}R_d\over R_d^2}\right)\!-\F_{(\sigma)}\right]\!,
\ee
where $g_{(\sigma)}$ is the string frame metric with signature $(-1,+\dots,+)$, $\cR_{(\sigma)}$ is the associated Ricci curvature, and $\phi$ is the dilaton in $d$ dimensions. Defining the Einstein frame metric as 
\be
g_{\mu\nu} =  e^{ -{4 \over d - 2} \phi } g_{(\sigma)\mu\nu},
\label{fra}
\ee
all dimensionful quantities acquire a dilaton dressing. In Einstein frame, the temperature, supersymmetry breaking scale and free energy density are therefore
\be
T= {e^{{2\over d-2} \phi}\over 2\pi R_0},\qquad M= {e^{{2\over d-2} \phi}\over 2\pi R_9}={e^{\sqrt{d-1\over d-2}\Phi}\over 2\pi}, \qquad \F= e^{{2d\over d-2} \phi}\F_{(\sigma)}=T^d f(z,\eta,\zeta).
\label{SE}
\ee
Note that we have introduced a new field $\Phi$, the so-called ``no-scale modulus''~\cite{noscale}. In fact, defining 
\be
\Phi=\sqrt{\frac{d-2}{d-1}}\left(\frac{2\phi}{d-2}-\eta\right)\!,\qquad
\Phi_\bot=\frac{1}{\sqrt{d-1}}\left(2\phi+\eta\right)\!,
\label{defPhi}
\ee
the action takes a suitable form in terms of canonical fields,
\be
\S=\int\dd^{d}x\sqrt{-g}\left[\frac{\cR}{2}-{1\over 2}\partial_\mu\Phi\partial^\mu\Phi-{1\over 2}\partial_\mu\Phi_\bot\partial^\mu\Phi_\bot- {1\over 2}\partial_\mu\zeta\partial^\mu\zeta-\F\right]\!.
\ee

Interested in flat, homogeneous and isotropic cosmological evolutions, we consider a Friedmann-Lema\^itre-Roberstson-Walker metric and space-independent scalar fields,
\begin{equation}
\dd s^{2}=-\beta(x^0)^2(\dd x^0)^{2}+a(x^0)^{2}\sum_{i=1}^{d-1}(\dd x^{i})^{2},\quad \Phi(x^0),\quad \Phi_\bot(x^0),\quad \zeta(x^0),
\end{equation}
where the lapse function $\beta$ is found by analytic continuation of the Euclidean background. Hence, it is the circumference of $S^1(R_0)$ measured in Einstein frame, which is nothing but the inverse temperature,
\be
\beta=e^{-{2\over d-2} \phi}2\pi R_0= {1\over T}.
\ee
Friedmann equations can be found by varying the action with respect to $\beta$ and $a$. They can be rewritten in terms of the more conventional cosmic time defined by $\dd t=\beta \dd x^0$:
\begin{align}
&\frac{(d-1)(d-2)}{2}\, H^{2}=\K+\rho,\label{Fried1}\\
&\frac{(d-1)(d-2)}{2}\, H^{2}+(d-2)\dot{H}=-\K-P,\label{Fried2}
\end{align}
where dot-derivatives are with respect to $t$ and $H=\dot a/a$. In the above equations, $\K$ is the kinetic energy of the scalars, while $\rho$ and $P$ are the energy density and pressure arising from the 1-loop contribution $\F$,
\begin{equation}
\K=\frac{1}{2}\!\left(\dot{\Phi}^{2}+\dot{\Phi}_{\bot}^{2}+\dot{\zeta}^{2}\right),\qquad
\rho=\F-T\frac{\partial \F}{\partial T}, \qquad P=-\F.
\end{equation}
Notice that the variational principle we have used matches perfectly with the thermodynamics laws, 
\be
\rho={1\over V}\left({\partial (\beta F)\over \partial \beta}\right)_{V},\quad P=-\left({\partial F\over \partial V}\right)_\beta,\quad \where\quad V=(2\pi a)^{d-1}, \;\;F=V\F.
\ee
For convenience, we may write the thermal energy density and pressure as
\be
\rho=T^{d}r(z,\eta,\zeta), \quad P=T^d p(z,\eta,\zeta),\quad \where\quad
r=f_z-(d-1)f,\quad p=-f,
\ee
and  $f_x=\partial f/\partial x$, for $x=z,\eta,\zeta$. With these notations, the scalar-field equations of motion take the form,
\begin{align}
\ddot \Phi+(d-1)H\dot \Phi &=-{\partial \F\over \partial \Phi}=-T^d \left(\sqrt{d-1\over d-2}\, f_z-\sqrt{d-2\over d-1}\, f_\eta\right) \!,\label{Phieq}\\
\ddot \Phi_\bot+(d-1)H\dot \Phi_\bot&= -{\partial \F\over \partial \Phi_\bot}=-{T^d\over \sqrt{d-1}}\, f_\eta ,\label{boteq}\\
\ddot \zeta+(d-1)H\dot \zeta&=-{\partial \F\over \partial \zeta}=-T^d f_\zeta.\label{zeq}
\end{align}

Combining the equations, it can be seen that Eq.~(\ref{Fried2}) can be replaced by an equation that can be solved~\cite{cosmo_phases},
\be
\label{aTcst}
{\dot \rho + \dot P\over \rho+P} +(d-1)H = {\dot T \over T} \quad \Longrightarrow\quad (aT)^{d-1}\big(r(z,\eta,\zeta)+p(z,\eta,\zeta)\big)=S,
\ee
where $S$ is the integration constant. The latter can be interpreted as the entropy of the universe since the above result implies
\be
U-TS=-PV\equiv F, \quad \where \quad U=V\rho.
\ee 


\subsection{Radiation-like dominated solutions and stabilization}
\label{RS}

When the supplementary massless states at the self-dual point contain more bosons than fermions, $\nFt-\nBt<0$ (recall that this means $\epsilon$ even in our example as defined in Eq.~(\ref{background})), the thermal effective potential $\F$ admits a local minimum at $\zeta=0$ (see Fig.~\ref{well}). As shown in Ref.~\cite{cosmo_phases}, this can yield a dynamical stabilization of $\zeta$ at the origin. In this subsection, we review these results, since they will be used later on to infer the behavior of the more involved  mechanism of mass generation for dark matter.  


\paragraph{Stabilization at the bottom of the well:}

Let us first describe a particular cosmological solution. Clearly, $\zeta\equiv 0$ solves Eq.~(\ref{zeq}). Since $f(z,\eta,0)$ is independent of $\eta$, Eq.~(\ref{boteq}) is satisfied for an arbitrary constant $\Phi_\bot\equiv \Phi_{\bot 0}$. It turns out to be convenient to replace Eq.~(\ref{Phieq}) by a differential equation for $z$. The latter involves a potential for $z$, which admits a minimum at some critical point $\tilde z_{\rm c}$ if and only if the massless spectrum of the model satisfies
\be
0<{N_{\rm F}-N_{\rm B}\over N_{\rm F}+N_{\rm B}}<\frac{1}{2^{d}-1}.
\label{rangeN}
\ee
In that case, $z\equiv \tilde z_{\rm c}$ is a solution,  where $\tilde z_{\rm c}$ is the unique root of the equation
\be
\tilde r(\tilde z_{\rm c})=d\tilde p(\tilde z_{\rm c}),\quad \where \quad \tilde r(z)\equiv r(z,\eta,0),\quad\tilde p(z)\equiv p(z,\eta,0)>0.
\label{rp}
\ee 
Note that this corresponds to the state equation of radiation in $d+1$ dimensions, $\rho=d\times P$.\footnote{With $n$ internal circles involved in the Scherk-Schwarz breaking of supersymmetry, this generalizes to $\rho=(d-1+n)P$~\cite{R4R5}} When the model-dependent quantity $(N_{\rm F}-N_{\rm B})/( N_{\rm F}+N_{\rm B})$ varies from 0 to its upper bound, $\tilde z_{\rm c}$ varies from $+\infty$ to $-\infty$.\footnote{\label{nb}We may include the lower bound 0 in Eq.~(\ref{rangeN}), at which we formally have $\tilde z_{\rm c}=+\infty$. In that case,  $z(t)$ is actually  running away rather than being stabilized at some finite value. The class of theories satisfying $N_{\rm F}=N_{\rm B}$ and sometimes referred to as ``super no-scale models'' may be of particular interest~\cite{Abel:2015oxa,Kounnas:2016gmz,Itoyama:1986ei,CFP}.} We are left with the Friedmann equation~(\ref{Fried1}), which turns out to take  the form $H^2=\tilde C_{\rm r}/a^{d}$. Ultimately, we find a critical solution
 \be
 \begin{aligned}
  &\zeta\equiv 0,\quad \Phi_\bot\equiv \Phi_{\bot 0},\quad M(t)\equiv T(t)\times e^{\tilde z_{\rm c}}\equiv {1\over a(t)} \times e^{\tilde z_{\rm c}}\!\left({S\over \tilde r(\tilde z_{\rm c})+\tilde p(\tilde z_{\rm c})}\right)^{1\over d-1},\\
  \where \quad & a(t)=t^{{2\over d}}\times \Big({d\over 2}\sqrt{\tilde C_{\rm r}}\Big)^{2\over d},\quad \tilde C_{\rm r}={2(d-1)\over d(d-2)^2}\,\tilde r(\tilde z_{\rm c})\!\left({S\over \tilde r(\tilde z_{\rm c})+\tilde p(\tilde z_{\rm c})}\right)^{d\over d-1}>0.
\end{aligned}
\label{RL}
\ee
To summarize, the supersymmetry breaking scale and the temperature evolve proportionally to the inverse of the scale factor. Moreover, this solution is compatible with weak string coupling. This can be seen by using Eq.~(\ref{defPhi}) to derive the time-dependence of the dilaton,
\be
e^{2{d-1\over d-2}\phi(t)}=2\pi M(t)\, e^{\sqrt{d-1}\Phi_{\bot0}},
\ee
which decreases with time. 

Note that along this very peculiar trajectory, $H^2\propto T^d$ as if the universe was filled with pure radiation in $d$ dimensions, which seems in contradiction with the result $\rho=d\times P$. In fact, using Friedmann equation~(\ref{Fried1}), the puzzle is resolved by observing that the {\em total} energy density and pressure satisfy
\be
\left.\begin{matrix}
\rho_{\rm tot}=\dis {1\over 2}\dot\Phi^2+\rho={(d-1)^2\over d(d-2)}\, \rho\\
\, P_{\rm tot}=\dis {1\over 2}\dot\Phi^2+P={d-1\over d(d-2)}\, \rho
\end{matrix}\right\}\quad \Longrightarrow\quad \rho_{\rm tot}=(d-1)P_{\rm tot}.
\ee
In other words, the classical kinetic energy of the no-scale modulus combines with the thermal free energy of the infinite towers of KK modes along $X^9$, to yield a ``radiation-like''  cosmological evolution \ie indistinguishable with that of a universe only filled with thermalized massless states.  

The local stability of this solution against small fluctuations has been shown analytically in Ref.~\cite{cosmo_phases}.\footnote{See also Refs~\cite{Liu:2011nw}, for supersymmetric theories at finite temperature.} Hence, for arbitrary initial conditions close enough to the trajectory of Eq.~(\ref{RL}), the generic evolution is attracted to the critical one. For this reason, we refer to these generic cosmological evolutions as ``radiation-like dominated solutions''. When converging to $\tilde z_{\rm c}$, $z(t)$ may or may not oscillate, depending on the initial conditions. Moreover, $\zeta$ always undergoes damped oscillations and eventually stabilizes at 0. Notice that this is a remarkable effect. In the literature, when such a scalar field has a constant mass and oscillates in a well, its energy does not dilute fast enough when the universe expands, the scalar does not stabilize, and the universe is not entering in a radiation-dominated era. To bypass this fact, known as the ``cosmological moduli problem'', the decay of the massive scalar field is invoked, which can lead to new difficulties such as an excessive entropy production~\cite{cosmomodprob}. 

The cosmological moduli problem does not occur in our string theory framework because the mass $m$ (measured in Einstein frame) of $\zeta$ is not constant. From Eq.~(\ref{mas}), we find
\be
m^2={2\over \pi}\, e^{{4\over d-2}\phi}\, T^{d-2},
\ee
which drops with time and increases the damping of the oscillations of $\zeta$, which is not anymore solely due to the friction resulting from the expansion of the universe. Eventually,  the energy stored in the modulus dilutes faster than the radiation-like density $\rho_{\rm tot}$ (or its component associated with the $N_{\rm F}+N_{\rm B}$ true species of radiation), so that $\zeta$ stabilizes. 

For completeness, we point out that when $(N_{\rm F}-N_{\rm B})/( N_{\rm F}+N_{\rm B})>1/(2^d-1)$, the supersymmetry breaking direction $X^9$ spontaneously decompactifies, and the supersymmetry breaking is screened by thermal effects. The generic evolution is naturally interpreted in a $(d+1)$-dimensional anisotropic universe, which is radiation dominated~\cite{cosmo_phases2}. Our purpose being eventually to describe the destabilization of $\zeta$ arising when supersymmetry breaking effects dominate over thermal ones, this case in not interesting to us in the present work.  
Alternatively, when $(N_{\rm F}-N_{\rm B})/( N_{\rm F}+N_{\rm B})<0$, the initially expanding universe stops growing and then collapses, with domination of moduli kinetic energy~\cite{attractor}. These remarks justify why we restrict our models to satisfy Eq.~(\ref{rangeN}).


\paragraph{Freezing along the plateaus:} 

The above attractor mechanism is only local, in the sense that initial conditions too far from the critical solution with $\zeta=0$ may yield a different behavior. In particular, when $\zeta$ is along one of the  plateaus shown in Fig.~\ref{well}, the $\nFt+\nBt$ states are heavier than $T$ and $M$ and yield exponentially suppressed contributions. Neglecting these terms, we have
\be
f(z,\eta,\zeta)=-(n_{\rm F}+n_{\rm B})f^{(d)}_{\rm T}(z)+(n_{\rm F}-n_{\rm B})f^{(d)}_{\rm V}(z).
\ee

Let us first describe a new critical solution. Clearly, the equations of motion~(\ref{zeq}) and~(\ref{boteq})  of $\zeta$ and $\Phi_\bot$ are satisfied when these fields are arbitrary constants $\zeta_0$ and $\Phi_{\bot 0}$. As explained in~\cite{cosmo_phases}, one can proceed as before and find that, provided that the model satisfies  
\be
0<{n_{\rm F}-n_{\rm B}\over n_{\rm F}+n_{\rm B}}<\frac{1}{2^{d}-1},
\label{rangen}
\ee
a particular solution exists with constant $z\equiv z_{\rm c}$, where $z_{\rm c}$ is the unique root of the equation\footnote{As in Footnote~\ref{nb}, we may include the lower bound 0 in Eq.~(\ref{rangen}), and have $z(t)$ running away towards $z_{\rm c}$ formally equal to $+\infty$.}
\be
\hat r(z_{\rm c})=d\hat p(z_{\rm c}),\quad \where \quad \hat r(z)\equiv r(z,\eta,0),\quad\hat p(z)\equiv p(z,\eta,0).
\ee 
Altogether, this peculiar evolution is
 \be
 \begin{aligned}
  &\zeta\equiv \zeta_0,\quad \Phi_\bot\equiv \Phi_{\bot 0},\quad M(t)\equiv T(t)\times e^{z_{\rm c}}\equiv {1\over a(t)} \times e^{z_{\rm c}}\!\left({S\over \hat r( z_{\rm c})+\hat p(z_{\rm c})}\right)^{1\over d-1},\\
  \where \quad & a(t)=t^{{2\over d}}\times \Big({d\over 2}\sqrt{\hat C_{\rm r}}\Big)^{2\over d},\quad \hat C_{\rm r}={2(d-1)\over d(d-2)^2}\,\hat r(z_{\rm c})\!\left({S\over \hat r(z_{\rm c})+\hat p(z_{\rm c})}\right)^{d\over d-1}>0,
\end{aligned}
\label{RLp}
\ee
which is radiation-like. It is also stable against small fluctuations. In other words, initial conditions close enough to the trajectory~(\ref{RLp}) yield evolutions attracted to the critical one, and are therefore radiation-like dominated. 


\paragraph{No eternal fall out of the plateaus: } 

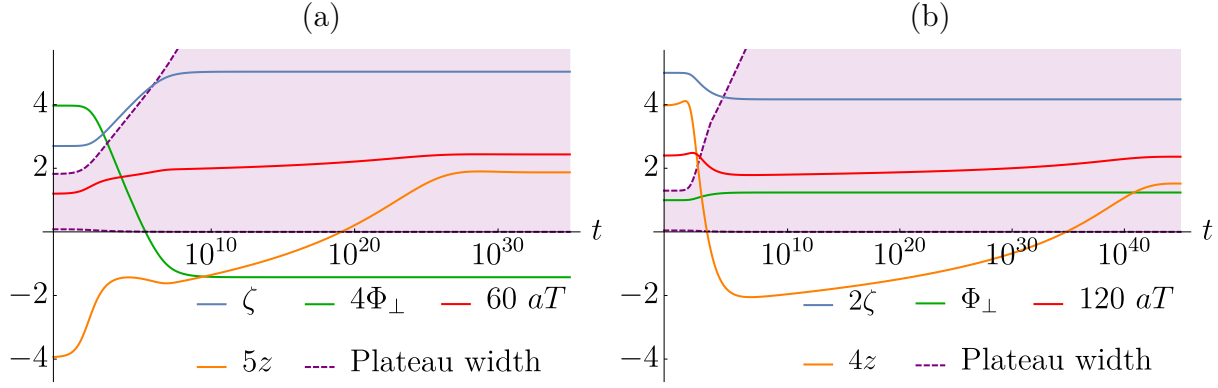
\begin{figure}[!h]
\centering
\vspace{0.7cm}
\includegraphics[scale=0.32]{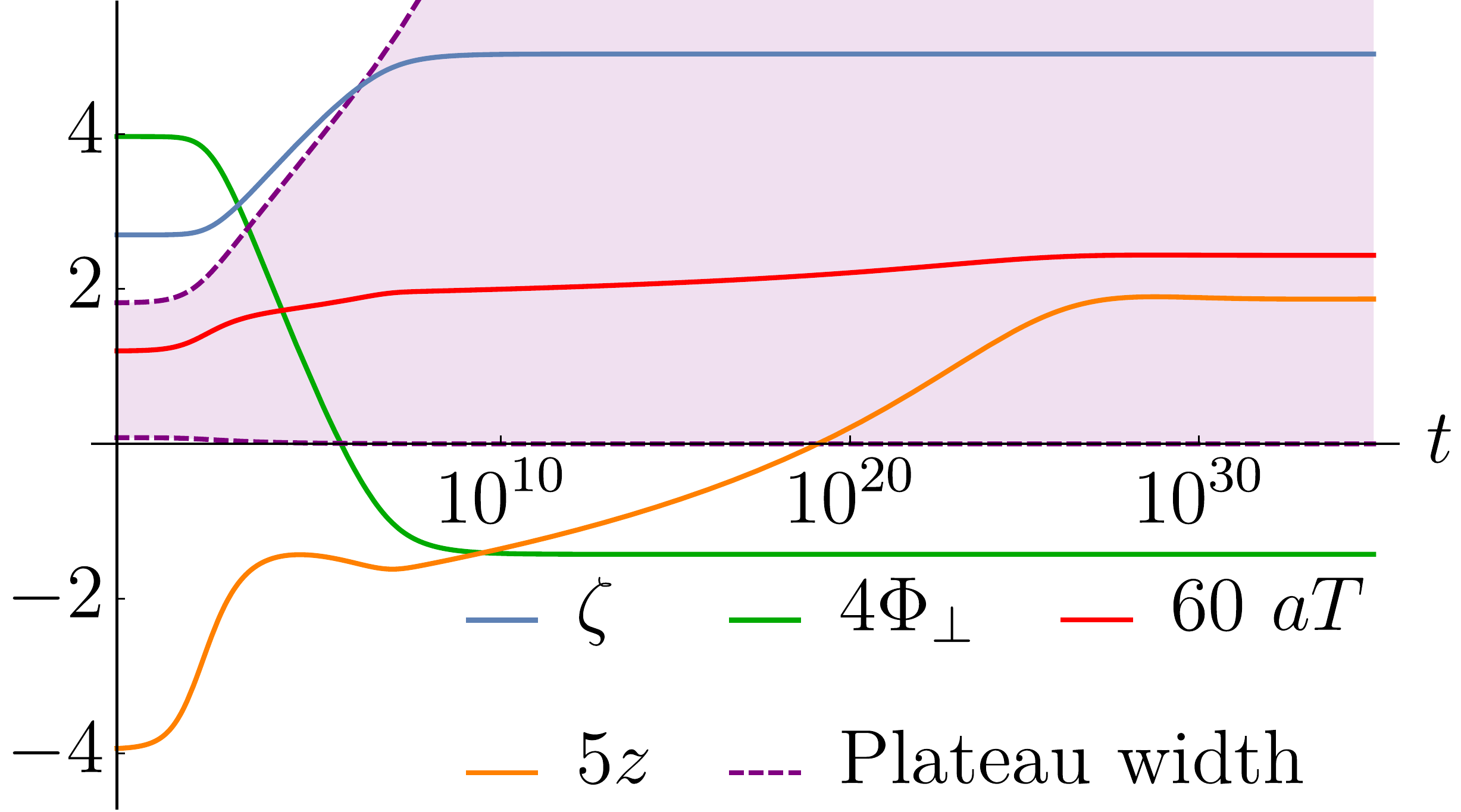}
\includegraphics[scale=0.32]{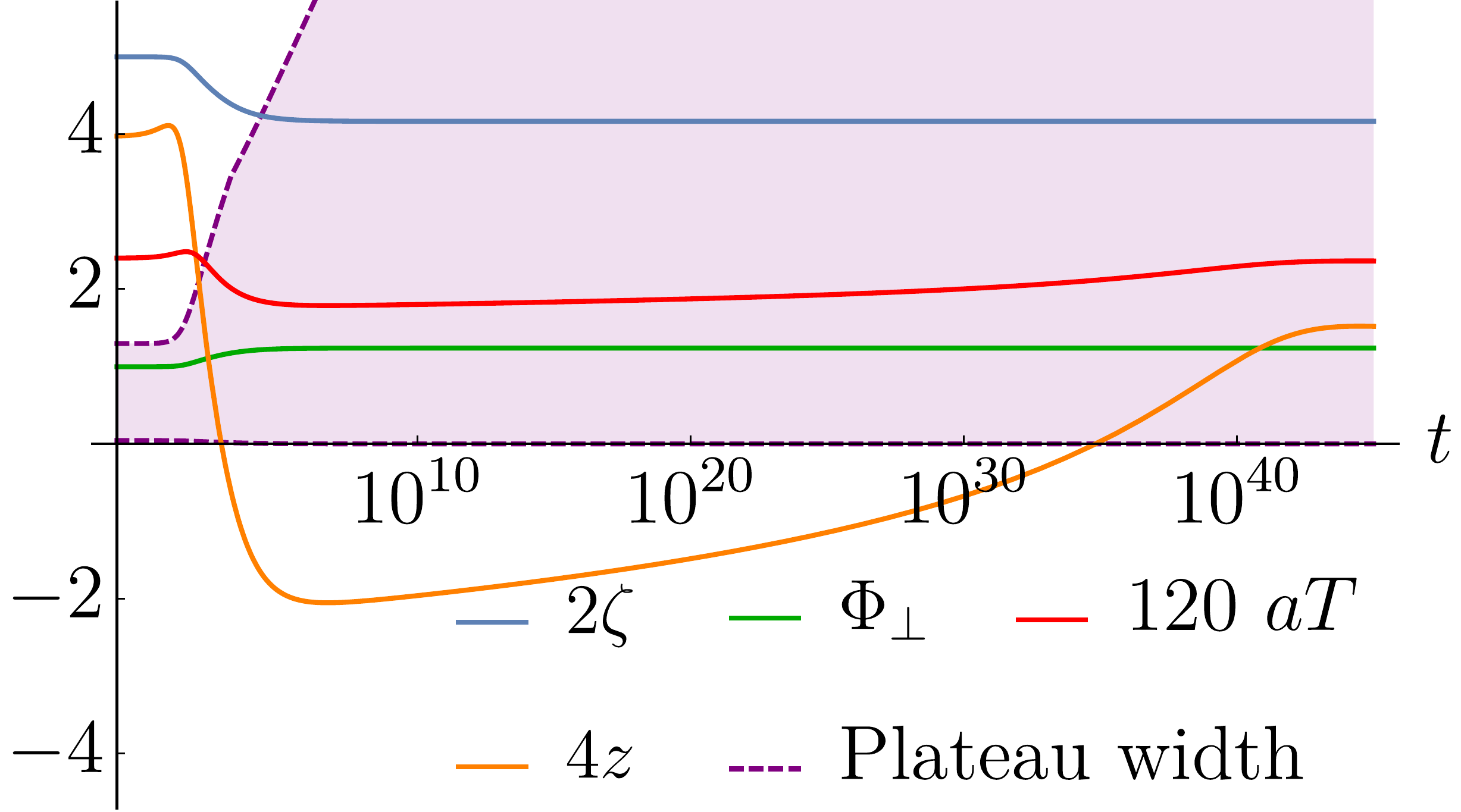}
\begin{picture}(0,0)
\put(-350,135){(a)}
\put(-120,135){(b)}
\end{picture}
\caption{\footnotesize (a)~{\em  Numerical simulation of $\zeta(t)$, $z(t)$, $\Phi_\bot(t)$ and $a(t)T(t)$, when $\zeta$ sits initially in the decreasing exponential part of its potential (see Figs~\ref{well} and \ref{bump}). The purple dashed curve shows the width of the plateau which increases with time. $\zeta(t)$ begins by increasing while falling along the potential until it is caught by the growing plateau. It then stabilizes while $z(t)$, $\Phi_\bot(t)$ and $a(t)T(t)$ eventually reach their asymptotic values.} (b)~{\em Same simulation when the exponential region of the thermal potential increases with $\zeta$, at initial time. $\zeta(t)$ begins by decreasing, thus approaching the plateau where it eventually freezes.}}
\label{plaC}
\end{figure}

For completeness, even if we will not make use of this property in the dark-matter generation mechanism described in the following section, we mention that the attraction towards the flat regions of the thermal effective potential of $\zeta$ is even stronger than may be expected. When the exponential behavior of $\F_{(\sigma)}$ is decreasing, as shown in Fig.~\ref{well} and~\ref{bump}, we may worry about the possibility that $|\zeta|$ falls out of the plateaus when it rolls enough to exceed $\min(\ln R_0 ,\ln R_9)$. In that case, one may think that the internal direction $X^d$  may spontaneously decompactify. However,  analytic arguments in favor of an attraction of $|\zeta|$ back to the plateaus was raised in Ref.~\cite{cosmo_phases,cosmo_phases2}. To figure out what is going on, we can simulate numerically the system of differential equations when $\zeta$ is initially located on the waterfall part of the potential, with low enough velocity. The evolutions in 4 dimensions of $\zeta(t)$, $z(t)$, $\Phi_\bot(t)$ and the product $a(t)T(t)$ are plotted for a set of generic initial conditions in Fig.~\ref{plaC}a.
While it is not a surprise to see $|\zeta(t)|$ to increase, the boundary  $ \min(\ln R_0(t),\ln R_9(t))$ of the plateau (delimited by the shaded area) increases faster and eventually catches up $\zeta(t)$. When the latter is back to the plateau, the evolution is attracted as before towards the critical solution of Eq.~({\ref{RLp}). 

When $\nF-\nB >0$ while $e^z$ is large enough, the exponential behavior of $\F_{(\sigma)}$ as a function of $\zeta$ is increasing. When this is the case, the attraction back to the plateau is naturally expected to be even more efficient than in the above waterfall case. As shown in Fig.~\ref{plaC}b, this expectation turns out to be confirmed by numerical simulation. As seen on the plots, $|\zeta(t)|$ starts by decreasing and then freezes once it is caught by the plateau. Notice however, that as long as $z(t)>0$ holds, the width of the plateaus is given  by $\ln R_9(t)$, while when  $z(t)<0$ it is determined by $\ln R_0(t)$. As a result, a change of the  slope of the time-dependent length of the plateaus is observed when $z(t)\simeq 0$.   


\subsection{Dynamical phase transition and mass generation} 
\label{transi}

We are now ready to describe the mechanism that triggers the phase transition responsible for generating large masses for states that will be interpreted as dark-matter constituents in the next section. The key point is to have an excess of massless fermionic modes at the self-dual point, $\nFt-\nBt>0$ ($\epsilon$ odd  in our example).


\paragraph{Qualitative expectations and specification of the models:} 

In order to infer what the mechanism will turn to lead to, let us remind what we learned from Eq.~(\ref{mas}). When $T$ is sufficiently larger than $M$ (\ie $e^z$ is small enough), $\zeta$ is massive. Assuming $\zeta$ to be initially in its potential well (see Fig.~\ref{well}), provided that Eq.~(\ref{rangeN}) holds, we expect the generic cosmological evolution to approach the critical solution of Eq.~(\ref{RL}). This attraction may be definitive, if $e^{\tilde z_{\rm c}}$ is low enough for maintaining  $\zeta$ massive throughout  the convergence of the evolution  towards the critical one. In that case, the behavior of the universe is identical to that described for $\nFt-\nBt<0$, with a stabilization of $\zeta$ at the origin.  

However, in models such that  $e^{\tilde z_{\rm c}}$ is large enough for $\zeta$ to be tachyonic, the above attraction of $z(t)$ towards $\tilde z_{\rm c}$ forces the squared mass of $\zeta$ to change sign during the evolution. The critical solution still exists, but  becomes unstable at this stage. In fact, the potential well of $\zeta$ becomes the bump shown in Fig.~\ref{bump} and a Higgs-like transition is expected to  occur, responsible for the destabilization of $\zeta$ away from the origin. The latter slides along the bump until it reaches one of the plateaus. Once there, assuming that Eq.~(\ref{rangen}) is satisfied, we have shown in the previous section that $\zeta$ gets frozen, due to the friction arising from the expansion of the universe. The final behavior of the evolution is thus attracted to the critical solution, Eq.~(\ref{RLp}), and is therefore radiation-like dominated.  

To summarize, for the mechanism to take place, the following conditions must hold:
\be
\begin{aligned}
(i)\quad&\nFt-\nBt>0&\!\!\!\!\!\!\!\!\!\!\!\!\!\!\! \!\!\!\!\!\!\!\!\!\!\!\!\!\!\!\!\!\!\!\!\!\!\!\!\!\!\!\!\!\!\!\!\!\!\!\mbox{(more extra massless fermions than bosons at $\zeta=0$),} \\
(ii)\quad&0<{n_{\rm F}-n_{\rm B}\over n_{\rm F}+n_{\rm B}}\;\;\and\;\;  {N_{\rm F}-N_{\rm B}\over N_{\rm F}+N_{\rm B}}<\frac{1}{2^{d}-1}&\mbox{($z_{\rm c}$ and $\tilde z_{\rm c}$ exist),}\\
(iii)\quad&f^{(d-2)}_{\rm T}(\tilde z_{\rm c})<{\nFt-\nBt\over  \nFt+ \nBt}\, f^{(d-2)}_{\rm V}(\tilde z_{\rm c})&\mbox{($\zeta$ tachyonic at $z=\tilde z_{\rm c}).$}
\end{aligned}
\ee
In our examples, these constraints translate  into
\be
(i)\;\;\epsilon \mbox{ odd}, \qquad (ii)\;\; 0<8-d_{\rm en}\;\;\and\;\; {10-d_{\rm en}\over 506+d_{\rm en}}<{1\over 2^d-1}, \qquad (iii)\;\; \tilde z_c>0,
\ee
which admit solutions in various dimensions:  

$\bullet$ For $d=3$,  we can have $d_{\rm en}=0,2,4,6$. The limit case $d_{\rm en}=8$ can be included if we allow $\tilde z_{\rm c}$ to be $+\infty$. The $d_{\rm en}$ roots of $G_{\rm en}$ can be realized at $SU(2)$ and/or $SU(3)$ enhanced gauge symmetry points of the Narain lattice of the internal $T^5$.

$\bullet$ For $d=4$, $d_{\rm en}=0,2,4,6$ (and possibly 8) are allowed.  The $d_{\rm en}$ roots can be realized at $SU(2)$ and/or $SU(3)$ points of the Narain lattice of $T^4$.

$\bullet$ For $d=5$, only $d_{\rm en}=4,6$ (or 8) are allowed and realized at $SU(2)$ and/or $SU(3)$ points of the Narain lattice of $T^3$.

$\bullet$ There is no solution for $d=6$, 7 and 8 in our examples. 


\paragraph{Numerical simulations:} 

Unlike critical solutions that describe asymptotic behaviors, the phase transition is a transient regime. Thus, solving analytically the equations of motions to describe explicitly the associated  solutions for generic initial conditions is difficult. For this reason, we have simulated numerically  the system of differential equations.  The results match with all the qualitative expectations described above.

Our choice of initial conditions at $t=0$ is such that the universe expands ($\dot a(0)>0$), with the temperature $T(0)$ slightly higher than the supersymmetry breaking scale $M(0)$ ($z(0)\lesssim 0$). Moreover, $\zeta(0)$ is anywhere in its well, with low enough velocity. Notice that if we assume throughout this paper the temperature (and supersymmetry breaking scale)  to be lower than the Hagedorn temperature, $R_0>R_H$, naturalness invites us to choose $R_0(0)$ equal to few units (counted in $\sqrt{\alpha'}$).\footnote{This is at least the case if we imagine that the cosmological era we describe is occurring right after a Hagedorn era characterized by  a temperature $T_{(\sigma)}$ comparable to the string scale. Such an intrinsically stringy epoch may describe a change of string vacuum~\cite{AtickWitten}, or bouncing cosmologies~\cite{Kounnas:2011fk}, which are alternative to the big bang and inflationary scenarios.} Note that such a radius $R_0(0)$ is enough for neglecting the exponentially suppressed contributions to the free energy, as done in Sect.~\ref{S2}. Second, the well has an initial width which is not very small, say of order $1/10$, and no severe fine tuning is required for $\zeta(0)$ to sit inside it.

 As shown in Fig.~\ref{oscillations}a, 
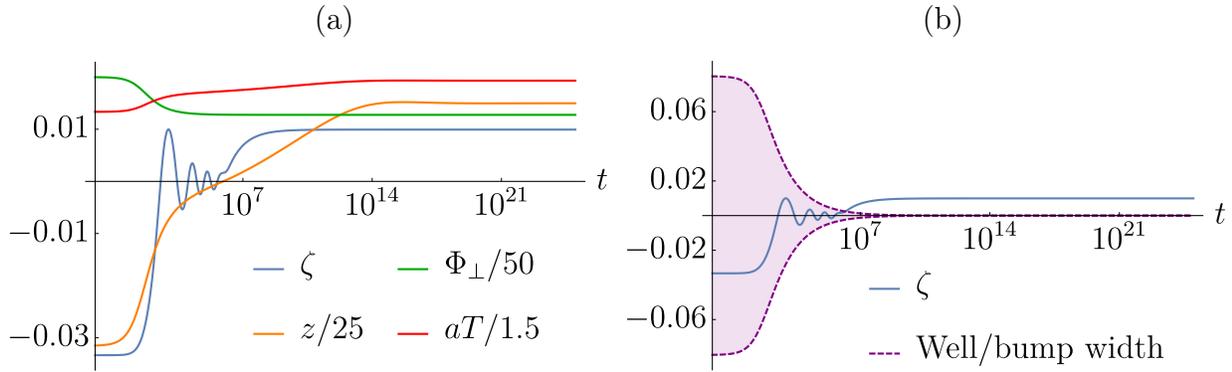
\begin{figure}[!h]
\centering
\vspace{0.5cm}
\includegraphics[scale=0.33]{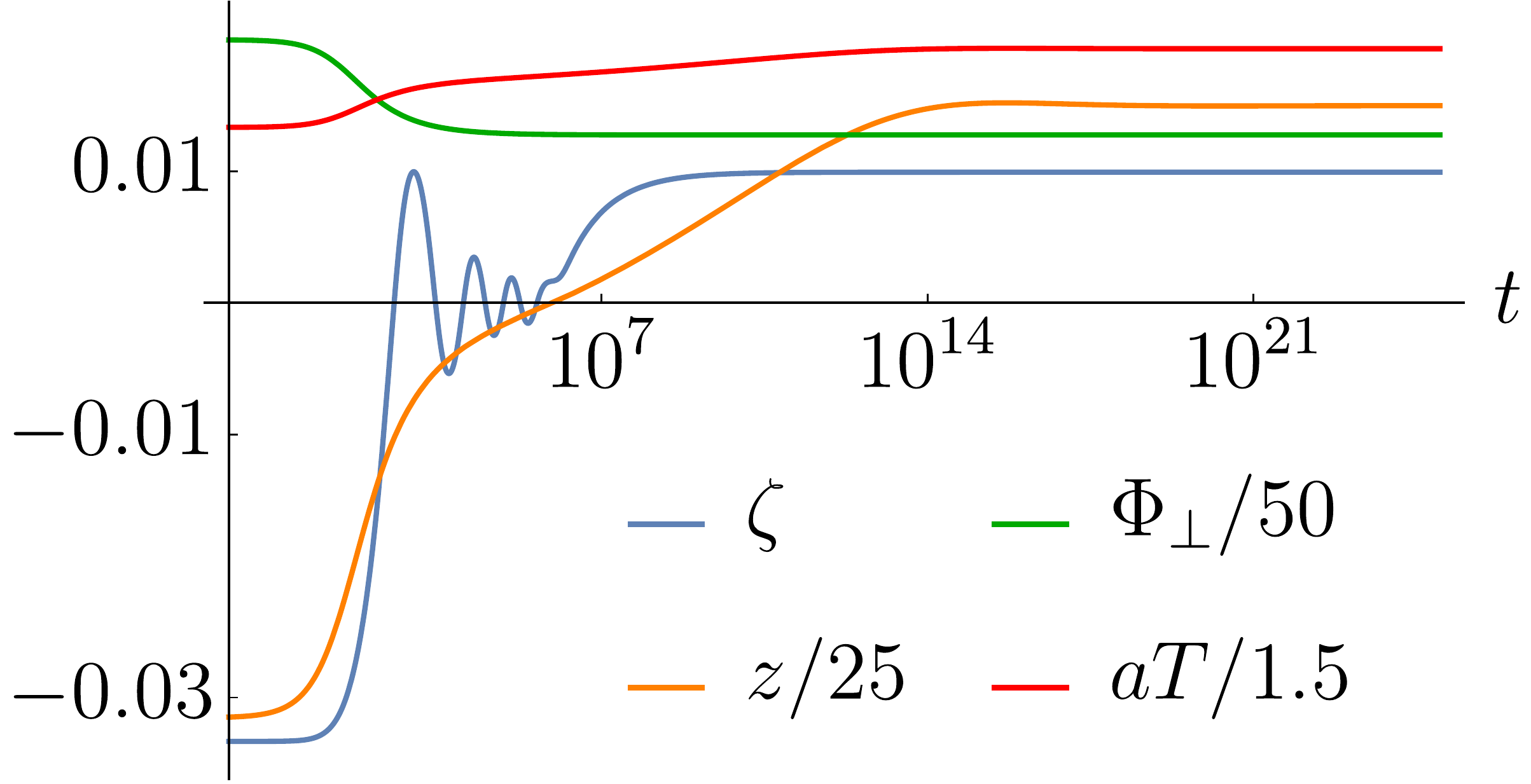}
\includegraphics[scale=0.33]{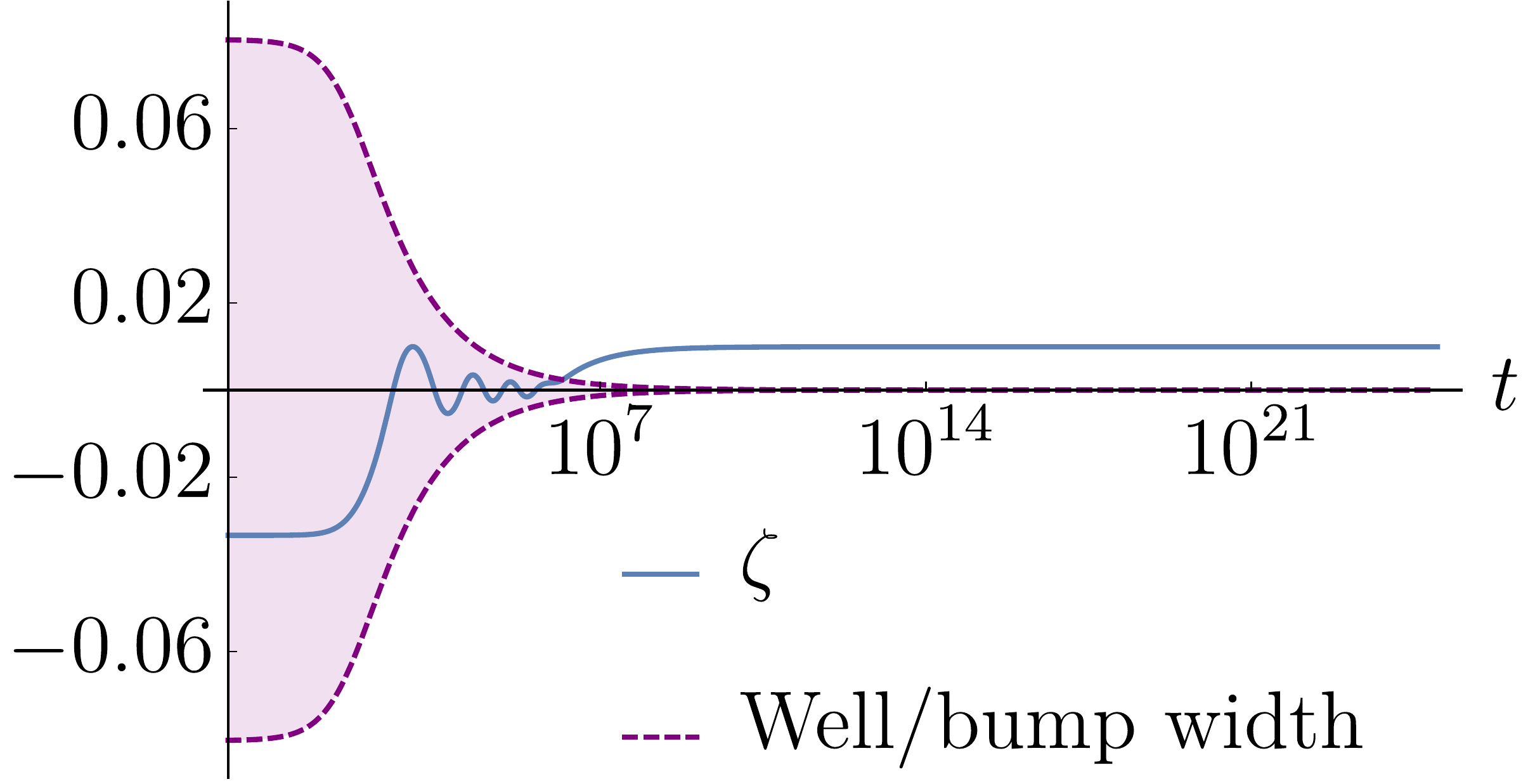}
\begin{picture}(0,0)
\put(-350,130){(a)}
\put(-120,130){(b)}
\end{picture}
\caption{\footnotesize (a)~{\em  Numerical simulation for $d=4$ of $\zeta(t)$, $z(t)$, $\Phi_\bot(t)$ and $a(t)T(t)$, in a model that realizes the dynamical phase transition responsible for a large mass generation of initially massless states. $\zeta(t)$ oscillates with damping around 0 as long as $z(t)<0$. When the latter become positive, $\zeta(t)$ condenses away from the origin.} (b)~{\em The shaded area represents the width of the well or bump. The oscillations are within the well, while freezing takes place away from the bump, where the potential for $\zeta$ is flat.}}
\label{oscillations}
\end{figure}
letting the system evolve in $d=4$ dimensions, $\zeta(t)$ starts oscillating with damping around the origin, while $z(t)$ is increasing (to approach $\tilde z_{\rm c}>0$). It is only when $z(t)$ becomes positive, so that the well turns into a bump, that $\zeta$, which is still close to~0, acquires some potential energy and eventually starts sliding along the hill before freezing. Meanwhile, $z(t)$ and $\Phi_\bot(t)$ converge to $z_{\rm c}>0$ and some constant $\Phi_{\bot0}$, while the product $a(t)T(t)$ also reaches a constant value. Notice that in Fig.~\ref{oscillations}a, because the final value of $|\zeta|$ is lower than $|\zeta(0)|$, one may think that the modulus remains stuck on the bump. However, the width of the hill decreases with time and eventually the motion of the modulus is coming to an end along a plateau. This can be seen on Fig.~\ref{oscillations}b, which shows the evolution of the width of the bump with time. To be specific, for the particular initial conditions we have chosen in the simulation, the final (string frame) mass $|R_d-1/R_d|\simeq 2|\zeta|$ of the $\nFt+\nBt$ states is of the order of $2\%$ of the string scale, which is rapidly several orders of magnitude larger than $T_{(\sigma)}$ and $M_{(\sigma)}$ that  keep on dropping. As a result, the $\nFt+\nBt$ modes yield exponentially suppressed contributions to the free energy, thus implying $\zeta$ to sit on a plateau. 

Before proceeding, we would like to stress that for the sake of simplicity, we have only allowed in our analytic and numerical analyzes a minimal set of moduli fields to vary. In particular, we could have treated the 4 degrees of freedom $(G+B)_{ij}$ in Eq.~\ref{background}) as dynamical, by generalizing the results of Ref.~\cite{CP} in presence of finite temperature. However, the mass generation mechanism we have presented would still take place. Moreover, it turns out that in our simple model, the $SO(16)\times SO(16)$ Wilson lines are  tachyonic for large enough $e^z$~\cite{Kounnas:2016gmz,CP}, and it could be artificial to maintain them static.\footnote{To figure out whether they can be destabilized or not, the ratio $M/T$ above which  they are tachyonic should be compared with $e^{z_c}$.} However, type I string models that satisfy the constraint $\nF-\nB>0$ and are tachyon free have been recently constructed~\cite{ADLP}, and the phase transition we have discussed may be implemented in their heterotic dual descriptions.


\section{Relic density evolution}
\label{dark}

The ratio of the mass induced by the phase transition to the temperature being large, the Boltzmann number density of the $\nFt+\nBt$ (with $\nFt-\nBt>0$ in generic models) initially massless degrees of freedom  may drastically decrease. In the present section, we explain how the expansion of the universe may nevertheless imply a non-trivial relic density of non-relativistic matter to survive.


\subsection{Dark-matter thermal freeze-out}

In the well-known {\em thermal} scenario of cosmology, the dark-matter  number density evolution throughout the universe history results from the competition between two opposite effects: On the one hand, number-depleting interactions between dark matter and the Standard Model\footnote{\label{TT} Note that number-depleting interactions within the dark sector could also maintain a thermal equilibrium in the dark sector with its own temperature. In that case, all occurrences of ``Standard-Model particles'' in the coming text should be replaced by ``massless dark-matter particles''.  However, it is natural to assume gauge interactions between the dark sector and the visible sector if we deal with only one temperature.} (typically, through its annihilation cross-section $\sigma_{{\rm DM}\leftrightarrow {\rm SM}}$) give the possibility for dark-matter particles to constantly readjust their number density $n_{{\rm DM}}$ to its Boltzmann equilibrium value $n_{{\rm DM,eq}}$.  The processes by which this happens are  two-to-two, of the form ${\rm DM}+{\rm DM}\rightarrow {\rm SM}+{\rm SM}$. On the other hand, the expansion of the universe tends to make interactions between dark-matter and Standard-Model particles more unlikely to happen, since it lowers their respective number densities. Hence, such a dilution renders a thermal equilibrium between dark-matter and Standard-Model particles more difficult to maintain. 

Before presenting how our string theory framework provides an alternative  way to generate a non-relativistic component of the universe energy density, let us first review how a non-vanishing relic density of dark matter is  generated in the usual thermal scenario.


\paragraph{Cold dark matter scenario:} 

In the standard thermal scenario,\footnote{In standard cosmology, there is no dynamical dilaton field and the Einstein frame is always implicit.}  a dark-matter particle has a constant mass $m_{{\rm DM}}$, and interacts with the Standard Model through two-to-two processes, whose annihilation cross-section is denoted $\sigma_{{\rm DM}\leftrightarrow {\rm SM}}$. For visualizing the chronology of the dark-matter number density, we draw in Fig.~\ref{fig:thermalDM}
\begin{figure}[!h]
\begin{center}
\includegraphics[width=0.62\linewidth]{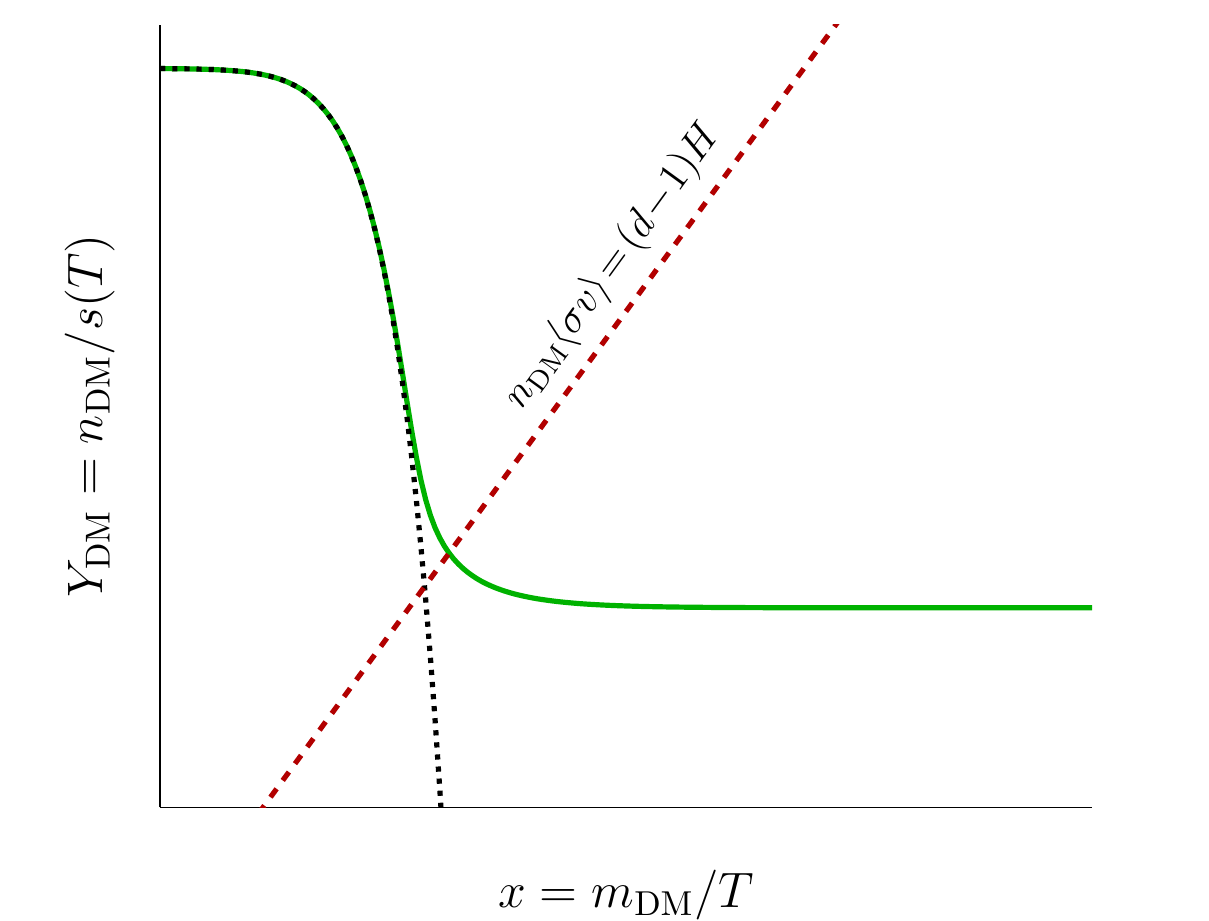}
\caption{\label{fig:thermalDM}\footnotesize Evolution of the dark-matter yield $Y_{{\rm DM}}=n_{{\rm DM}}/s$ as a function of $x=m_{\rm DM}/T$ in the standard thermal scenario, in 4 dimensions (green solid line). The black-dotted line represents the value that the yield would follow if thermal equilibrium could be maintained all along the history of the universe. Freeze-out takes place when interactions with the Standard-Model particles are too weak, as compared to the expansion rate of the universe (after crossing of the red dashed line).}
\end{center}
\end{figure}
the typical evolution of the so-called yield, 
\be
Y_{{\rm DM}}={n_{{\rm DM}}(T)\over s(T)},
\label{Y}
\ee 
where $s(T)\propto T^{d-1}\propto 1/a^{d-1}$ is the entropy density of the thermal bath. In this figure, the evolution is parametrized by $x=m_{\rm DM}/T$, and we  have chosen arbitrary values of the dark-matter mass and annihilation cross-section.  The thermal scenario of cold dark-matter production can be summarized as follows:

$\bullet$ $T\gg m_{{\rm DM}}$: At early times, both dark-matter and Standard-Model particles are relativistic. If their interactions are strong enough, dark matter is maintained chemically in thermal equilibrium with the Standard-Model bath so that  $n_{{\rm DM}}=n_{{\rm DM,eq}}\propto T^{d-1}$. The entropy density of the universe evolving as $s\propto T^{d-1}$, the dark-matter yield $Y_{\rm DM}=n/s$ is  initially constant, as can be seen on the left-hand side of Fig.~\ref{fig:thermalDM}. Quantitatively, the interaction is strong enough when $(d-1)H<n_{{\rm DM}}\langle\sigma_{{\rm DM} \leftrightarrow {\rm SM}}v\rangle$, where $v$ stands for the dark-matter particles relative velocities and the brackets $\langle\, \cdots\rangle$ denote the mean over velocity distribution. 

$\bullet$ $T\lesssim m_{{\rm DM}}$: When the temperature drops under the dark-matter mass, dark-matter particles become non-relativistic. The Boltzmann distribution becomes exponentially suppressed, $n_{{\rm DM,eq}}\sim e^{-m/T}$, and dark-matter particles tend to annihilate more and more into Standard-Model particles in order to maintain equilibrium, therefore lowering their number density (see Fig.~\ref{fig:thermalDM}, where the black-dotted and green-solid lines drop together). Standard-Model particles which have energy $\langle E\rangle\sim T$ are less and less able to produce them back. Again, such number depletion is possible as long as interactions are strong enough \ie the condition $(d-1)H<n_{{\rm DM}}\langle\sigma_{{\rm DM} \leftrightarrow {\rm SM}}v\rangle$ is still satisfied.

$\bullet$ $T\lesssim m_{{\rm DM}}$  and $(d-1)H\gtrsim n_{{\rm DM}}\langle\sigma_{{\rm DM} \leftrightarrow {\rm SM}}v\rangle$: After dark-matter particles started annihilating significantly into Standard Model particles, the expansion rate of the universe can dominate over the annihilation rate, which leads to a {\em chemical} decoupling. In Fig.~\ref{fig:thermalDM}, this corresponds to the point where the red dashed curve is crossed. The universe being radiation dominated before chemical decoupling, this curve turns out to satisfy  $Y_{\rm DM}\propto x^{{d\over 2}-1}$, which is linear for $d=4$. 
At the crossing, the annihilation of dark matter stops, dark-matter particles {\em freeze-out}, and a relic density of dark-matter particles remains as a non-relativistic component of the universe. Thus, the number density evolves again as $n_{{\rm DM}}\propto 1/a^{d-1}\propto T^{d-1}$ and the yield becomes constant. 

Formally speaking, such a freeze-out can be described by the Boltzmann equation, in terms of the dark-matter number density $n_{{\rm DM}}(t)$,
\begin{equation}\label{eq:boltzmann}
\frac{{\rm d}n_{{\rm DM}}}{{\rm d}t}+(d-1)H n_{{\rm DM}}=-\langle\sigma_{{\rm DM}\leftrightarrow {\rm SM}}v\rangle\!\left[n_{{\rm DM}}^2-n_{{\rm DM,eq}}^2\right]\!.
\end{equation}
In this formulation, Standard-Model particles are assumed to be in thermal equilibrium. Moreover, the dark-matter particles number density at equilibrium is defined as
 \be
 \label{eq:neq}
 n_{{\rm DM,eq}}(t)=\tilde n  \int {{\rm d}^{d-1}\vec{k} \over (2\pi)^{d-1}}\, {1\over e^{{\sqrt{m_{\rm DM}^2+\vec{k}^2}\over T(t)}}\pm1},
 \ee
 where $\tilde n$ is the number of  dark-matter degrees of freedom, with either Bose-Einstein of Fermi-Dirac statistics and the integration runs over the momentum $\vec{k}$ of the particles.  
Neglecting the expansion rate $(d-1)H n_{{\rm DM}}$ in Eq.~\eqref{eq:boltzmann}, an over-density (under-density) of dark-matter particles as compared to the equilibrium value would pull (push) the dark-matter density  back to its equilibrium value. Therefore, as long as the expansion term can be neglected as compared to the interaction term in the right-hand side of Eq.~\eqref{eq:boltzmann}, the dark-matter density follows its equilibrium value,
\begin{equation}
n_{{\rm DM}}\simeq n_{{\rm DM,eq}} \qquad \text{(before freeze-out).}
\end{equation}
Conversely, when the expansion starts dominating over interactions in the equation, one can neglect the right-hand side and obtain
\begin{equation}
n_{{\rm DM}}\propto {1\over a^{d-1}} \qquad \text{(after freeze-out).}
\end{equation}


\paragraph{The neutrino case:} 

So far, we have been discussing the case of a cold dark matter, decoupling from the thermal bath after it becomes non-relativistic. This guarantees that the dark matter is cold enough  for not streaming freely on large distances after it is produced, ensuring that the large scale structures are preserved, in agreement with present cosmological measurements~\cite{Viel:2005qj}.

Nevertheless, the condition that a particle becomes non-relativistic before it decouples from the thermal bath is not necessary. In fact, as long as its interaction with Standard-Model particles becomes weak enough at a temperature larger than the dark-matter mass, dark-matter particles can still be relativistic when they freeze-out. This is exactly what happens to neutrinos, which decouple at a temperature $T\sim \mathrm{MeV}$ from the thermal bath, far before they become non-relativistic. This mechanism will take place in some circumstances in the string theory framework we are now turning to.  


\paragraph{Our string theory scenario:}

As mentioned above, experimental constraints on structure formation impose that dark matter constitutes today a large, non-relativistic component of the energy density.

Moreover, we have seen that the key point for a cold dark-matter scenario to be successful is to have at some point the temperature lower than the dark-matter mass, and to ensure that its interaction with radiation is weak enough, for a significant amount of dark-matter particles to remain frozen after they decouple. In string theory, gauge interactions between the dark and visible sectors may or may not exist. In the examples we have constructed in Sect.~\ref{S2}, our main motivation was to present the simplest realization of a phase transition responsible for a mass generation of initially massless states. Being maximally supersymmetric (in a spontaneously broken version, \eg $\N=4\to \N=0$ in 4 dimensions), massless matter cannot be chiral and  there is no Standard-Model to discuss in this context. However, models compatible with chirality~\cite{Faraggi:2017cnh} and realizing an $\N=1\to \N=0$ spontaneous breaking of supersymmetry in four dimensions may be considered. For instance, they can be realized \via orbifold compactifications or fermionic constructions. Implementing the mass generation mechanism in such models, dark and Standard-Model sectors may for instance be unified prior to the phase transition in a  gauge theory based on $E_6$, $SO(10)$, a Pati-Salam gauge group, etc. In such a case, a significant annihilation cross section $\sigma_{\rm DM\leftrightarrow SM}$ is then natural. In the following, we assume the string theory model to be realistic enough for such a non-trivial cross-section to exist. Other possibilities may however be considered, as noticed in Footnote~\ref{TT}.

In the standard thermal scenario, before freeze-out, when the universe is radiation dominated, as well as after dark matter decouples, the (approximate) relations $s\propto T^{d-1}\propto 1/a^{d-1}$ we have used extensively hold. Instead, the cosmological evolution derived from string theory before freeze-out is radiation-like dominated and satisfies Eq.~(\ref{aTcst}). To be specific, prior to the phase transition, $\zeta$ oscillates in the well, and the evolution approaches the critical, radiation-like solution~(\ref{RL}). Similarly, if the destabilization process ends and $\zeta$ is stuck on a plateau, the evolution is attracted towards the second critical, radiation-like evolution, Eq.~(\ref{RLp}). Hence, before freeze-out, the (approximate) relations  $s\propto T^{d-1}\propto 1/a^{d-1}$ hold and the yield definition in Eq.~(\ref{Y}) can be used as in the standard thermal scenario. However, after dark matter decouples from the thermal bath and eventually dominates, $z(t)$ and $\Phi_\bot(t)$ have no more reason to be static, implying $S/(aT)^{d-1}$ not to be a constant. Consequently, we will use in this regime an alternative definition of the yield,
\be
Y_{\rm DM}\propto n_{\rm DM}\, a^{d-1},
\label{Y2}
\ee
which clearly matches with Eq.~(\ref{Y}) before freeze-out.  

The main difference of our scenario with the usual thermal case is that the dark-matter particles masses are driven by the value of $\zeta=\ln R_d$ and suddenly increase after the phase transition described in Sec.~\ref{transi} takes place. Since we have shown that the transition is sufficient to render part of the spectrum spontaneously non-relativistic, such particles can freeze-out and constitute a dark component of the universe. To describe qualitatively this mechanism, we consider in the following the limit case where the condensation of $\zeta$, \ie the mass jump, is much faster than all other processes, such as the evolutions of the temperature and number density $n_{\rm DM}$. Hence, we assume from now on that while the temperature drops, $\nFt$ and $\nBt$ (with $\nFt-\nBt>0$) dark-matter fermionic and bosonic degrees of freedom are massless before the phase transition, and acquire ``instantaneously'' a mass $m_{{\rm DM}}$  at a temperature $T=T_{\rm c}$ (both measured in Einstein frame)
\begin{equation}
m(T)=\left\{\begin{array}{c l}
0 &\quad \text{for $T>T_{\rm c}$,}\\
m_{{\rm DM}} &\quad \text{for $T<T_{\rm c}$.}
\end{array}\right.
\end{equation}
Notice that $T_{\rm c}$ is not determined {\em a priori}. Our assumption of instantaneity supposes $\zeta(t)$ starts sliding from the top of the hill when $z(t)\simeq z_{\rm c}$ (its final value after the transition) but this condition fixes only the ratio $M/T$ to its critical value $e^{z_{\rm c}}=M_{\rm c}/T_{\rm c}$ at the transition. 
It turns out that depending on the ratio $x_{\rm c}=m_{\rm DM}/T_{\rm c}$, two qualitatively different situations may occur, as can be seen in Fig.~\ref{relic}, which represents the evolution of the yield. For completeness, a third case is also shown on this figure.  
To describe them, we define the dark-matter number densities right before and right after the transition as $n_{{\rm DM,eq}}^{0}$ and $n_{{\rm DM,eq}}^{m_{\rm DM}}$, respectively. Moreover, we  treat $\langle\sigma_{\rm DM\leftrightarrow SM} v\rangle$ as not varying at the transition. 

$\bullet$ {\bf Case 1:} At $T=T_{\rm c}$, \;\;$n_{{\rm DM,eq}}^{0}\langle\sigma_{\rm DM\leftrightarrow SM} v\rangle > (d-1)H$ \;\;and \;\;$n_{{\rm DM,eq}}^{m_{{\rm DM}}}\langle\sigma_{\rm DM\leftrightarrow SM} v\rangle >(d-1)H$\\
In Fig.~\ref{relic}, 
\begin{figure}[!h]
\begin{center}
\includegraphics[width=0.62\linewidth]{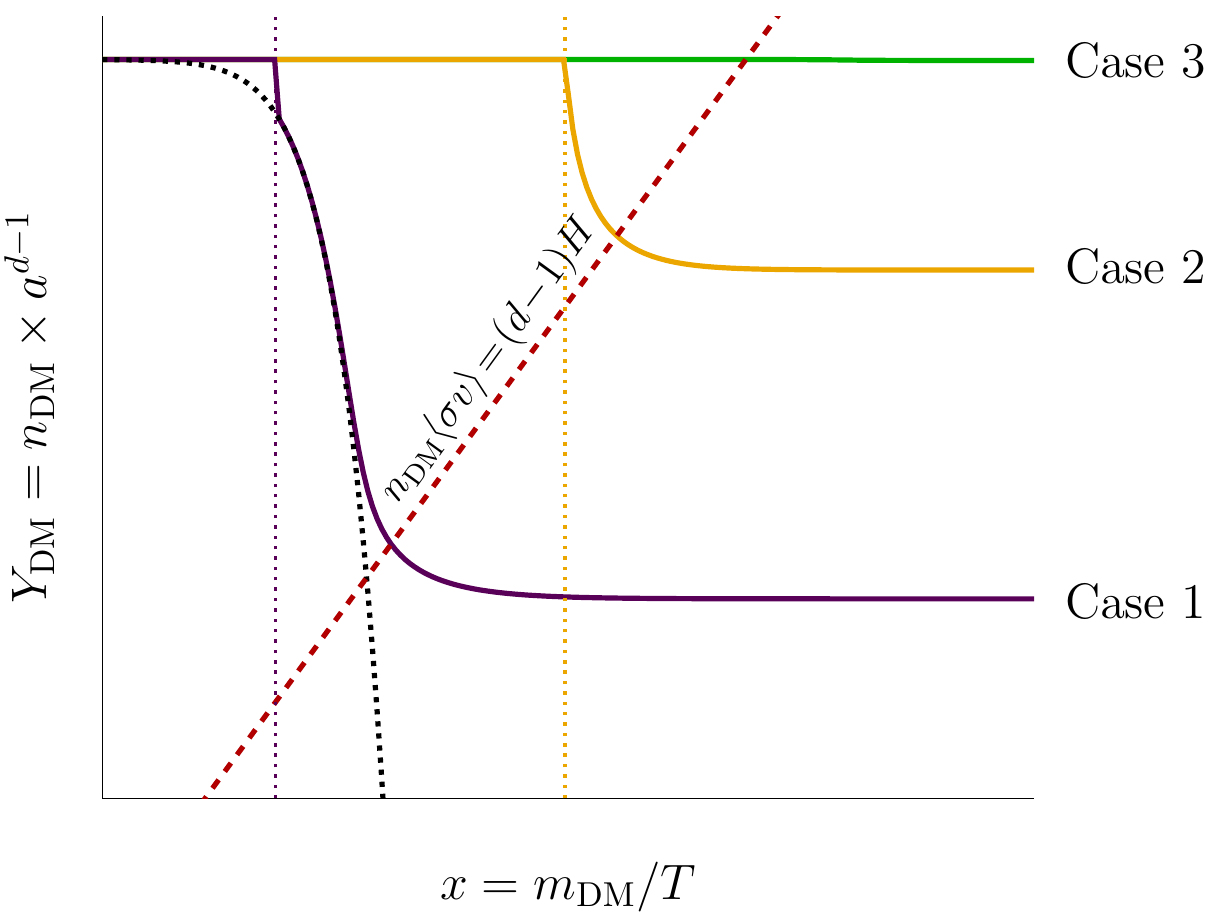}
\caption{\label{relic}\footnotesize Evolution of the dark-matter yield $Y_{\rm DM}=n_{\rm DM}a^{d-1}$ ($\propto n_{\rm DM}/s$ before freeze-out) as a function of $x=m_{\rm DM}/T$ in the string theory scenario, in 4 dimensions. The black-dotted line represents the value that the yield would follow if the dark-matter particles always had a constant mass equal to $m_{\rm DM}$, and if thermal equilibrium could be maintained all along the history of the universe. 
In Cases~1 (violet line) and~2 (yellow line), two different values of the phase transition temperature $T_{\rm c}$ are considered. In Case~3 (green line), the decoupling from the thermal bath takes place while dark matter is still relativistic, and no phase transition can take place thereafter. }
\end{center}
\end{figure}
this corresponds to the case where the black dotted curve  is above the red dashed curve at $x_{\rm c}=m_{\rm DM}/T_{\rm c}$: Before $x_{\rm c}$, the  dark-matter constituent being thermal radiation, the yield is constant.  After the dark matter acquires its mass, it can annihilate sufficiently for its number density to drop all the way down to its new equilibrium value (see the violet line in Fig.~\ref{relic}). Then, one recovers the case of a standard thermal cold dark-matter scenario. Chemical decoupling eventually occurs,  the density quits equilibrium and the yield of Eq.~(\ref{Y2}) freezes-out.

$\bullet$ {\bf Case 2:} At $T=T_{\rm c}$, \;\;$n_{{\rm DM,eq}}^{0}\langle\sigma_{\rm DM\leftrightarrow SM} v\rangle > (d-1)H$ \;\;and \;\;$n_{{\rm DM,eq}}^{m_{{\rm DM}}}\langle\sigma_{\rm DM\leftrightarrow SM} v\rangle <(d-1)H$\\
In Fig.~\ref{relic}, the black dotted curve  is below the red dashed curve at $x_{\rm c}=m_{\rm DM}/T_{\rm c}$: In this case, the massless dark matter acquires a mass while its number density is still sufficient for the annihilation process to be efficient for a while. However, while decreasing, the chemical decoupling limit is reached before a new thermal equilibrium can be established. Therefore, dark matter freezes-out at an intermediate relic density (see the yellow line in Fig.~\ref{relic}).

$\bullet$ {\bf Case 3:} Thermal decoupling may also occur while dark-matter particles are still massless. In Fig.~\ref{relic}, the red dashed curve intersects the green horizontal line, while $\zeta$ has not  been destabilized yet. Before decoupling, the number density follows the relativistic equilibrium value $n_{{\rm DM,eq}}\propto T^{d-1}$. Then, dark-matter decouples while still relativistic, similarly to the neutrino case. The particle number gets frozen and its density keeps evolving as $1/a^{d-1}$ due to the universe expansion. Therefore, the alternative definition of the yield, Eq.~(\ref{Y2}), remains constant. In fact, it turns out that no mass generation can take place after decoupling, and we recover a usual hot dark-matter scenario. To reach these conclusions, let us focus on the free energy density component associated with the $\nFt+\nBt$ massless states, accompanied with their  KK modes along the supersymmetry breaking direction~$X^9$. Before freeze-out, the result (in string frame) can be extracted from Eq.~(\ref{mas}), 
\be
\begin{aligned}
\F_{{\rm DM}(\sigma)}=T_{(\sigma)}^d\Big\{&\!-(\nFt+\nBt)f^{(d)}_{\rm T}(z)+(\nFt-\nBt)f^{(d)}_{\rm V}(z)\\
&\!+{\zeta^2\over \pi T_{(\sigma)}^2}\!\left[(\nFt+\nBt)f^{(d-2)}_{\rm T}(z)-(\nFt-\nBt)f^{(d-2)}_{\rm V}(z)\right]\!+\O(\zeta^4)\Big\},
\end{aligned}
\label{fdm}
\ee 
where $\zeta\simeq 0$ is massive. If at the decoupling from the thermal bath the dark-matter energy density $\rho_{{\rm DM}(\sigma)}^{\rm f}$ and pressure $P_{{\rm DM}(\sigma)}^{\rm f}=-\F_{{\rm DM}(\sigma)}^{\rm f}$ can be derived from the above formula, the corresponding expressions at later times are 
\be
\rho_{{\rm DM}(\sigma)}= \rho_{{\rm DM}(\sigma)}^{\rm f} \bigg({a_{(\sigma)}^{\rm f}\over a_{(\sigma)}}\bigg)^{d-1}, \quad P_{{\rm DM}(\sigma)}= P_{{\rm DM}(\sigma)}^{\rm f} \bigg({a_{(\sigma)}^{\rm f}\over a_{(\sigma)}}\bigg)^{d-1}.
\ee
In our notations, $a_{(\sigma)}^{{\rm f}}$ is the scale factor when dark matter freezes out, and the scaling rule of the energy density and pressure results from the dilution arising from the universe expansion. Therefore,  Friedmann equations~(\ref{Fried1}) and~(\ref{Fried2}) are affected. However, more important to us  is that the potential of $\zeta$ in this regime is nothing but
\be
\F_{{\rm DM}(\sigma)}= \F_{{\rm DM}(\sigma)}^{\rm f} \bigg({a_{(\sigma)}^{\rm f}\over a_{(\sigma)}}\bigg)^{d-1}.
\ee
Up to an irrelevant overall scaling, its shape is frozen to that given at the decoupling, which we know is of a well shape around $\zeta=0$. As a result, the possibility that any phase transition would be responsible for the mass generation of dark-matter particles is ruled out. In fact,   when decoupling occurs in the massive phase of $\zeta$, dark matter remains hot, no matter the sign of  $\nFt-\nBt$ is.  

Let us finally comment on the fact that, so far, we have assumed the phase transition to happen instantaneously, as compared to the time scale necessary for dark-matter particles to readjust their number density to its equilibrium value. Indeed, while the mass of dark matter varies with time, the shape of the equilibrium density as defined in Eq.~\eqref{eq:neq} also changes with time. If the mass variation is slow enough, dark-matter particles could adiabatically annihilate into Standard-Model particles (\ie with $n_{\rm DM}$ following its mass-dependent equilibrium value), modifying the temperature at which the chemical decoupling would happen, and therefore the final value of the relic density. A careful solving of Boltzmann equation, together with a precise computation of the annihilation cross-section would be necessary to describe correctly such a situation, which will be addressed in a more complete study of the phenomenon in the future.


\subsection{Dark-matter relic energy density}

Our aim is now to partially compute the relic energy density after phase transition and freeze-out, in Cases~1 and~2. In the examples of Sect.~\ref{S2}, the  dark-particles spectrum for $\epsilon$ odd amounts to $\nFt$ KK towers of fermionic degrees of freedom, together with their bosonic superpartners. The former have  KK momentum $m_9/R_9$ along the Scherk-Schwarz direction $X^9$, while the latter have shifted momentum $(m_9+{1\over 2})/R_9$.\footnote{This can be seen by taking the limit $R_0\to +\infty$ in Eq.~(\ref{free1}) and applying a Poisson summation over $\tilde k_9$.}

To proceed, we need to specify the velocity distribution of each dark-matter KK species after the phase transition. Since the lowest string-frame mass $|R_d-1/R_d|$ of the KK modes is already much higher than the temperature $T_{(\sigma)}$, we will simply assume a Dirac distribution of all velocities \ie that all dark-matter particles are at rest.  The total energy associated to the dark sector has then two origins. On the one hand, for each KK species, the mass has to be weighted by the particle number obtained when all particles have decoupled from the thermal bath (see the final constant values reached by the yield in Cases~1,~2 in Fig.~\ref{relic}). On the other hand, the vacuum energy of all degrees of freedom also contributes. The latter is the effective potential at zero-temperature associated to the KK towers of fermionic and bosonic modes. As can be seen from Sect.~\ref{S2}, such an energy is exponentially suppressed in $R_9|R_d-1/R_d|\gg 1$, since the mass of the zero-momentum mode along  $X^9$ is  much larger than the supersymmetry breaking scale $M_{(\sigma)}$.\footnote{In field theory, this vacuum energy is infinite for the bosonic modes alone, infinite for the fermionic modes alone, but their sum turns out to be finite for arbitrary $|R_d-1/R_d|$. Technically, this finiteness arises exactly as that  of the free energy at finite temperature evaluated for a supersymmetric spectrum. String theory yields the same final answer, up to contributions arising from stringy heavy modes not present in field theory (see Ref.~\cite{Casimir} for more details).} To be consistent with the fact that we have neglected throughout this paper all such exponentially suppressed heavy modes contributions, we restrict to the energy arising from the relic particles at rest. 

From the above considerations, the total relic dark-matter energy measured in string frame can be expressed in terms of the particle numbers and dynamical radii $R_d$, $R_9$ as follows\footnote{If $R_9$ is time-dependent, $\zeta=\ln R_d$ may evolve again once the universe is matter dominated.},
\be
\label{Es}
\begin{aligned}
E_{(\sigma)}=\nFt(a_{(\sigma)}^{{\rm f}})^{d-1}\int \dd^{d-1}\vec{k}_{(\sigma)}\sum_{m_9}\Bigg[ &\, \N_{{\rm B},m_9+{1\over 2}}^{\rm f}\big(\vec k_{(\sigma)}\big) \sqrt{\Big(R_{d}-\frac{1}{R_{d}}\Big)^{2}+\Big(\frac{m_{9}+\frac{1}{2}}{R_{9}}\Big)^{2}}    \\
&- \N_{{\rm F},m_9}^{\rm f}\big(\vec k_{(\sigma)}\big) \sqrt{\Big(R_{d}-\frac{1}{R_{d}}\Big)^{2}+\Big(\frac{m_{9}}{R_{9}}\Big)^{2}} \,\Bigg].
\end{aligned}
\ee
Of course, a precise computation of the cross-section $\sigma_{\rm DM\leftrightarrow SM}$ is required to determine when decoupling takes place. This is compulsory to derive the value of the scale factor $a_{(\sigma)}^{{\rm f}}$ at this time. Knowledge of $\sigma_{\rm DM\leftrightarrow SM}$  is also necessary to figure out the final value of the yield (see Fig.~\ref{relic}) \ie the distributions $\N_{{\rm B},m_9+{1\over 2}}^{\rm f}$, $\N_{{\rm F},m_9}^{\rm f}$.  However, computing the cross section and distributions goes beyond the scope of the present paper and we  leave this study for later works.

In the end, the dark-matter contribution to the action for a homogeneous and isotropic universe is 
\be
\S_{\rm DM}=-\int N_{(\sigma)}a_{(\sigma)}^{d-1}\dd x^d\, \rho_{\rm DM(\sigma)} =- \int N a^{d-1}\dd x^d\, \rho_{\rm DM},
\ee
where we have expressed the result in either string or Einstein frames, with arbitrary definition of time \ie generic lapse functions $N_{(\sigma)}$ and $N$, respectively. From Eq.~(\ref{fra}), the relic dark-matter  energy densities are 
\be
\rho_{\rm DM(\sigma)} = \nFt\,\mathscr{E}(\eta,\zeta)\, \Big({a^{\rm f}_{(\sigma)}\over a_{(\sigma)}}\Big)^{d-1},\qquad \rho_{\rm DM} = \nFt\,(a^{\rm f}_{(\sigma)})^{d-1}\,\mathscr{E}(\eta,\zeta)\,  {e^{{2\over d-2}\phi}\over a^{d-1}},
\ee
where we have defined 
\be
\label{Es2}
\begin{aligned}
 \mathscr{E}(\eta,\zeta)=\int {\dd^{d-1}\vec{k}_{(\sigma)}\over (2\pi)^{d-1}}\sum_{m_9}\Bigg[ &\, \N_{{\rm B},m_9+{1\over 2}}^{\rm f}\big(\vec k_{(\sigma)}\big) \sqrt{\Big(R_{d}-\frac{1}{R_{d}}\Big)^{2}+\Big(\frac{m_{9}+\frac{1}{2}}{R_{9}}\Big)^{2}}    \\
&- \N_{{\rm F},m_9}^{\rm f}\big(\vec k_{(\sigma)}\big) \sqrt{\Big(R_{d}-\frac{1}{R_{d}}\Big)^{2}+\Big(\frac{m_{9}}{R_{9}}\Big)^{2}} \,\Bigg].
\end{aligned}
\ee
Of course, varying $\S_{\rm DM}$ with respect to either of the scale factors, one derives trivial pressures for cold dark-matter,
\be
P_{\rm DM(\sigma)} =0,\qquad P_{\rm DM} =0.
\ee
However, in string frame, the dark-matter energy density sources the equations of motion for $\zeta=\ln R_d$ and $M_{(\sigma)}=1/(2\pi R_9)$. In the Einstein frame, it affects the dynamics of $\zeta$, the no-scale modulus $\Phi$ and $\Phi_\bot$, as follows from Eq.~(\ref{SE}) and the relation
\be
e^{2{d-1\over d-2}\phi(t)}=2\pi M(t)\, e^{\sqrt{d-1}\Phi_{\bot}}.
\ee


\section{Conclusions and perspectives}
\label{conclu}

The mechanism we have presented  for generating non-relativistic dark matter  may be relevant for describing an intermediate era of the  cosmological history of the universe. At earlier times, the standard scenario assumes the existence of a period of inflation followed by reheating. While the possibility of realizing this picture in a ultraviolet complete theory is not clear so far~\cite{landswamp,Obied:2018sgi}, other possibilities, inherently stringy by nature, have also been considered. Among these  proposals, various  pre-big bang scenarios~\cite{CosmoPheno,Kounnas:2011fk} have been analyzed, or Hagedorn phase transitions~\cite{AtickWitten} may take place. 

Whatever the very early eras look like, assuming that at some later time the universe is flat, homogeneous, isotropic and thermalized,  we have found that the mechanism that triggers the phase transition responsible for the dark matter mass  is preceded by a ``Radiation-like Dominated'' evolution, which is an attractor of the dynamics. This means that the motion of the supersymmetry breaking scale $M(t)$ together with the thermal energy density and pressure associated to KK towers of states conspire for the universe to evolve as if it was dominated by pure radiation. In this regime, the temperature $T(t)$ and the supersymmetry breaking scale $M(t)$ are of the same order of magnitude. However, when the dark-matter particles suddenly become massive and freeze-out, their energy density eventually dominates over radiation and a preliminary numerical analysis of the system seems to yield a rapid increase of the ratio $M(t)/T(t)$. Hence,  a large hierarchy $M\gg T$ is dynamically generated, as must be the case to account for the smallness of the cosmic microwave background temperature, as compared to the very large supersymmetry breaking scale.

For the matter domination to end once dark energy takes over, the motion of $M(t)$ should come to a halt. We let for future work the proposal of a mechanism responsible for the stabilization of $M$. However, models yielding an extremely small (and positive) cosmological constant should be very peculiar. It could be that they satisfy conditions similar, and actually stronger, than those considered in Refs~\cite{Abel:2015oxa,Kounnas:2016gmz,Itoyama:1986ei}, which have vanishing effective potential at 1-loop.



\section*{Acknowledgement}
 
We are grateful to Quentin Bonnefoy and Emilian Dudas  for fruitful discussions. The research activities of L.H. are supported  by the Department of Energy under Grant DE-FG02-13ER41976/DE-SC0009913 and by CNRS. L.H. would like to thank the CPHT of Ecole Polytechnique for hospitality.
The work of H.P. is partially supported by the Royal Society International Cost Share Award. 




\begin{thebibliography}{99}
 
 
\bibitem{landswamp}
  C.~Vafa,
  ``The string landscape and the swampland,''
  hep-th/0509212;\\
   H.~Ooguri and C.~Vafa,
  ``On the geometry of the string landscape and the swampland,''
  Nucl.\ Phys.\ B {\bf 766} (2007) 21
  [hep-th/0605264].
   
   
   \bibitem{Aghanim:2018eyx}
  N.~Aghanim {\it et al.} [Planck Collaboration],
  ``Planck 2018 results. VI. Cosmological parameters,''
  arXiv:1807.06209 [astro-ph.CO].

\bibitem{Baumann:2009ds}
  D.~Baumann,
  ``Inflation,''
  arXiv:0907.5424 [hep-th].
  
  
   \bibitem{Obied:2018sgi}
  G.~Obied, H.~Ooguri, L.~Spodyneiko and C.~Vafa,
  ``De Sitter space and the swampland,''
  arXiv:1806.08362 [hep-th];\\
  P.~Agrawal, G.~Obied, P.~J.~Steinhardt and C.~Vafa,
  ``On the cosmological implications of the string swampland,''
  Phys.\ Lett.\ B {\bf 784} (2018) 271
  [arXiv:1806.09718 [hep-th]];\\
   H.~Ooguri, E.~Palti, G.~Shiu and C.~Vafa,
  ``Distance and de Sitter conjectures on the swampland,''
  Phys.\ Lett.\ B {\bf 788} (2019) 180
  [arXiv:1810.05506 [hep-th]].
  

\bibitem{Dashko:2018dsw}
  A.~Dashko and R.~Dick,
  ``The shadow of dark matter as a shadow of string theory,''
  arXiv:1809.01089 [hep-ph];\\
  B.~S.~Acharya, S.~A.~R.~Ellis, G.~L.~Kane, B.~D.~Nelson and M.~Perry,
  ``Categorisation and detection of dark matter candidates from string/M-theory hidden sectors,''
  JHEP {\bf 1809} (2018) 130
  [arXiv:1707.04530 [hep-ph]];\\
  G.~Honecker and W.~Staessens,
  ``On axionic dark matter in type IIA string theory,''
  Fortsch.\ Phys.\  {\bf 62} (2014) 115
  [arXiv:1312.4517 [hep-th]];\\
  G.~Shiu, P.~Soler and F.~Ye,
  ``Milli-charged dark matter in quantum gravity and string theory,''
  Phys.\ Rev.\ Lett.\  {\bf 110} (2013) no.24,  241304
  [arXiv:1302.5471 [hep-th]];\\
  A.~Zanzi,
  ``Dilaton stabilization and composite dark matter in the string frame of heterotic-M-theory,''
  arXiv:1210.4615 [hep-th];\\
  H.~Kleinert,
  ``The purely geometric part of  ``dark matter'' -- A fresh playground for ``string theory'',''
  Electron.\ J.\ Theor.\ Phys.\  {\bf 9} (2012) no.26,  27
  [arXiv:1107.2610 [gr-qc]];\\
  B.~S.~Acharya, G.~Kane and E.~Kuflik,
  ``Bounds on scalar masses in theories of moduli stabilization,''
  Int.\ J.\ Mod.\ Phys.\ A {\bf 29} (2014) 1450073
  [arXiv:1006.3272 [hep-ph]];\\
  S.~F.~King and J.~P.~Roberts,
  ``Natural dark matter from type I string theory,''
  JHEP {\bf 0701} (2007) 024
  [hep-ph/0608135];\\
  D.~Bailin, G.~V.~Kraniotis and A.~Love,
  ``Sparticle spectrum and dark matter in type I string theory with an intermediate scale,''
  Phys.\ Lett.\ B {\bf 491} (2000) 161
  [hep-ph/0007206];\\
  %
  K.~Benakli, J.~R.~Ellis and D.~V.~Nanopoulos,
  ``Natural candidates for superheavy dark matter in string and M-theory,''
  Phys.\ Rev.\ D {\bf 59} (1999) 047301
  [hep-ph/9803333].  


\bibitem{DDM}
  K.~R.~Dienes and B.~Thomas,
  ``Dynamical dark matter: II. An explicit model,''
  Phys.\ Rev.\ D {\bf 85} (2012) 083524
  [arXiv:1107.0721 [hep-ph]];\\
  K.~R.~Dienes and B.~Thomas,
  ``Dynamical dark matter: I. Theoretical overview,''
  Phys.\ Rev.\ D {\bf 85} (2012) 083523
  [arXiv:1106.4546 [hep-ph]]; \\
  K.~R.~Dienes, F.~Huang, S.~Su and B.~Thomas,
  ``Dynamical dark matter from strongly-coupled dark sectors,''
  Phys.\ Rev.\ D {\bf 95} (2017) no.4,  043526
  [arXiv:1610.04112 [hep-ph]].
  
  \bibitem{Dienes:2017zjq}
  K.~R.~Dienes, J.~Fennick, J.~Kumar and B.~Thomas,
  ``Dynamical dark matter from thermal freeze-out,''
  Phys.\ Rev.\ D {\bf 97} (2018) no.6,  063522
  [arXiv:1712.09919 [hep-ph]].

  \bibitem{Franca:2004kk}
  U.~Franca and R.~Rosenfeld,
  ``Variable-mass dark matter and the age of the Universe,''
  astro-ph/0412413.
  
  \bibitem{SSstring}
  R.~Rohm,
  ``Spontaneous supersymmetry breaking in supersymmetric string theories,''
  Nucl.\ Phys.\ B {\bf 237} (1984) 553;\\
  C.~Kounnas and M.~Porrati,
  ``Spontaneous supersymmetry breaking in string theory,''
  Nucl.\ Phys.\ B {\bf 310} (1988) 355;\\
  S.~Ferrara, C.~Kounnas and M.~Porrati,
  ``Superstring solutions with spontaneously broken four-dimensional supersymmetry,''
  Nucl.\ Phys.\  B {\bf 304} (1988) 500; \\
  S.~Ferrara, C.~Kounnas, M.~Porrati and F.~Zwirner,
  ``Superstrings with spontaneously broken supersymmetry and their effective theories,''
  Nucl.\ Phys.\ B {\bf 318} (1989) 75.
  
  \bibitem{Kounnas-Rostand}
  C.~Kounnas and B.~Rostand,
  ``Coordinate-dependent compactifications and discrete symmetries,''
  Nucl.\ Phys.\ B {\bf 341} (1990) 641.

  \bibitem{SS}
  J.~Scherk and J.~H.~Schwarz,
  ``Spontaneous breaking of supersymmetry through dimensional reduction,''
  Phys.\ Lett.\  B {\bf 82} (1979) 60;\\
%
  J.~Scherk and J.~H.~Schwarz,
 ``How to get masses from extra dimensions,''
  Nucl.\ Phys.\ B {\bf 153} (1979) 61.
  
  
\bibitem{attractor}
  F.~Bourliot, C.~Kounnas and H.~Partouche,
  ``Attraction to a radiation-like era in early superstring cosmologies,''
  Nucl.\ Phys.\ B {\bf 816} (2009) 227
  [arXiv:0902.1892 [hep-th]].

\bibitem{solcri}
  T.~Catelin-Jullien, C.~Kounnas, H.~Partouche and N.~Toumbas,
  ``Thermal/quantum effects and induced superstring cosmologies,''
  Nucl.\ Phys.\ B {\bf 797} (2008) 137
  [arXiv:0710.3895 [hep-th]].
 
 \bibitem{R4R5}
  T.~Catelin-Jullien, C.~Kounnas, H.~Partouche and N.~Toumbas,
  ``Induced superstring cosmologies and moduli stabilization,''
  Nucl.\ Phys.\ B {\bf 820} (2009) 290
  [arXiv:0901.0259 [hep-th]].
  
\bibitem{cosmo_phases}
  F.~Bourliot, J.~Estes, C.~Kounnas and H.~Partouche,
  ``Cosmological phases of the string thermal sffective potential,''
  Nucl.\ Phys.\ B {\bf 830} (2010) 330
  [arXiv:0908.1881 [hep-th]].
  
  \bibitem{cosmo_phases2}
  J.~Estes, C.~Kounnas and H.~Partouche,
  ``Superstring cosmology for $N_4 = 1 \to 0$ superstring vacua,''
  Fortsch.\ Phys.\  {\bf 59} (2011) 861
  [arXiv:1003.0471 [hep-th]].
  
\bibitem{GV}
  P.~H.~Ginsparg and C.~Vafa,
  ``Toroidal compactification of nonsupersymmetric heterotic strings,''
  Nucl.\ Phys.\ B {\bf 289} (1987) 414.
  
    \bibitem{cosmomodprob}
  B.~de Carlos, J.~A.~Casas, F.~Quevedo and E.~Roulet,
  ``Model independent properties and cosmological implications of the dilaton and moduli sectors of 4-d strings,''
  Phys.\ Lett.\  B {\bf 318} (1993) 447
  [arXiv:hep-ph/9308325];

  G.~D.~Coughlan, R.~Holman, P.~Ramond and G.~G.~Ross,
  ``Supersymmetry and the entropy crisis,''
  Phys.\ Lett.\  B {\bf 140}, 44 (1984).
  
  
\bibitem{KiritsisBook}
  E.~Kiritsis,
  ``String theory in a nutshell,''
  Princeton University Press, 2007.
  
\bibitem{o16}
  L.~Alvarez-Gaume, P.~H.~Ginsparg, G.~W.~Moore and C.~Vafa,
  ``An $O(16) \times O(16)$ heterotic string,''
  Phys.\ Lett.\ B {\bf 171} (1986) 155.
  
\bibitem{Kounnas:2016gmz}
  C.~Kounnas and H.~Partouche,
  ``Super no-scale models in string theory,''
  Nucl.\ Phys.\ B {\bf 913} (2016) 593
  [arXiv:1607.01767 [hep-th]];\\
%
  C.~Kounnas and H.~Partouche,
  ``$\N=2 \to 0$ super no-scale models and moduli quantum stability,''
  Nucl.\ Phys.\ B {\bf 919} (2017) 41
  [arXiv:1701.00545 [hep-th]].
  
    \bibitem{CP}
  T.~Coudarchet and H.~Partouche,
  ``Quantum no-scale regimes and moduli dynamics,''
  Nucl.\ Phys.\ B {\bf 933} (2018) 134
  [arXiv:1804.00466 [hep-th]].
  
\bibitem{Abel:2015oxa}
  S.~Abel, K.~R.~Dienes and E.~Mavroudi,
  ``Towards a nonsupersymmetric string phenomenology,''
  Phys.\ Rev.\ D {\bf 91} (2015) no.12,  126014
  [arXiv:1502.03087 [hep-th]].
  
   \bibitem{noscale}
  E.~Cremmer, S.~Ferrara, C.~Kounnas and D.~V.~Nanopoulos,
  ``Naturally vanishing cosmological constant in $\N=1$ supergravity,''
  Phys.\ Lett.\  B {\bf 133} (1983) 61;\\
  J.~R.~Ellis, C.~Kounnas and D.~V.~Nanopoulos,
  ``Phenomenological $SU(1,1)$ supergravity,''
  Nucl.\ Phys.\  B {\bf 241} (1984) 406;\\
  J.~R.~Ellis, A.~B.~Lahanas, D.~V.~Nanopoulos and K.~Tamvakis,
 ``No-scale supersymmetric standard model,''
  Phys.\ Lett.\  B {\bf 134} (1984) 429;\\
  J.~R.~Ellis, C.~Kounnas and D.~V.~Nanopoulos,
  ``No scale supersymmetric GUTs,''
  Nucl.\ Phys.\  B {\bf 247} (1984) 373.

\bibitem{Itoyama:1986ei}
  H.~Itoyama and T.~R.~Taylor,
  ``Supersymmetry restoration in the compactified $O(16) \times O(16)'$ heterotic string theory,''
  Phys.\ Lett.\ B {\bf 186} (1987) 129;\\
  %
S.~Abel, K.~R.~Dienes and E.~Mavroudi,
  ``GUT precursors and entwined SUSY: The phenomenology of stable nonsupersymmetric strings,''
  Phys.\ Rev.\ D {\bf 97} (2018) no.12,  126017
  [arXiv:1712.06894 [hep-ph]];\\
  S.~Abel and R.~J.~Stewart,
  ``On exponential suppression of the cosmological constant in non-SUSY strings at two loops and beyond,''
  Phys.\ Rev.\ D {\bf 96} (2017) 106013
  [arXiv:1701.06629 [hep-th]];\\
  I.~Florakis and J.~Rizos,
  ``Chiral heterotic strings with positive cosmological constant,''
  Nucl.\ Phys.\ B {\bf 913} (2016) 495
  [arXiv:1608.04582 [hep-th]].
  
  \bibitem{CFP}
  T.~Coudarchet, C.~Fleming and H.~Partouche,
  ``Quantum no-scale regimes in string theory,''
  Nucl.\ Phys.\ B {\bf 930} (2018) 235
  [arXiv:1711.09122 [hep-th]];\\
  H.~Partouche,
  ``Quantum no-scale regimes and string moduli,''
  Universe {\bf 4} (2018) no.11,  123
  [arXiv:1809.03572 [hep-th]].
  
\bibitem{Liu:2011nw}
  L.~Liu and H.~Partouche,
  ``Moduli stabilization in type II Calabi-Yau compactifications at finite temperature,''
  JHEP {\bf 1211} (2012) 079
  [arXiv:1111.7307 [hep-th]];\\
  J.~Estes, L.~Liu and H.~Partouche,
  ``Massless D-strings and moduli stabilization in type I cosmology,''
  JHEP {\bf 1106} (2011) 060
  [arXiv:1102.5001 [hep-th]].
  
  

  \bibitem{AtickWitten}
  J.~Atick and E.~Witten,
  ``The Hagedorn transition and the number of degrees of freedom of string
  theory,''
  Nucl.\ Phys.\  B {\bf 310}, 291 (1988);\\
%
  I.~Antoniadis and C.~Kounnas,
  ``Superstring phase transition at high temperature,''
  Phys.\ Lett.\  B {\bf 261} (1991) 369;\\
I.~Antoniadis, J.~P.~Derendinger and C.~Kounnas,
  ``Nonperturbative temperature instabilities in $\N=4$ strings,''
  Nucl.\ Phys.\ B {\bf 551} (1999) 41
  [hep-th/9902032];\\
  C.~Angelantonj, C.~Kounnas, H.~Partouche and N.~Toumbas,
  ``Resolution of Hagedorn singularity in superstrings with gravito-magnetic fluxes,''
  Nucl.\ Phys.\ B {\bf 809} (2009) 291
  [arXiv:0808.1357 [hep-th]].




\bibitem{Kounnas:2011fk}
  C.~Kounnas, H.~Partouche and N.~Toumbas,
  ``Thermal duality and non-singular cosmology in $d$-dimensional superstrings,''
  Nucl.\ Phys.\ B {\bf 855} (2012) 280
  [arXiv:1106.0946 [hep-th]];\\
  C.~Kounnas, H.~Partouche and N.~Toumbas,
  ``S-brane to thermal non-singular string cosmology,''
  Class.\ Quant.\ Grav.\  {\bf 29} (2012) 095014
  [arXiv:1111.5816 [hep-th]].

\bibitem{ADLP}
S.~Abel, E.~Dudas, D.~Lewis and H.~Partouche, ``Stability and vacuum energy in open string models with broken supersymmetry,'' CPHT-RR117.122018, DCPT-18/37, IPPP/18/112.


\bibitem{Viel:2005qj}
  M.~Viel, J.~Lesgourgues, M.~G.~Haehnelt, S.~Matarrese and A.~Riotto,
  ``Constraining warm dark matter candidates including sterile neutrinos and light gravitinos with WMAP and the Lyman-alpha forest,''
  Phys.\ Rev.\ D {\bf 71} (2005) 063534
  [astro-ph/0501562].

\bibitem{Faraggi:2017cnh}
  A.~E.~Faraggi, J.~Rizos and H.~Sonmez,
  ``Classification of Standard-like Heterotic-String Vacua,''
  Nucl.\ Phys.\ B {\bf 927} (2018) 1
  [arXiv:1709.08229 [hep-th]];\\
  B.~Assel, K.~Christodoulides, A.~E.~Faraggi, C.~Kounnas and J.~Rizos,
  ``Classification of heterotic Pati-Salam models,''
  arXiv:1007.2268 [hep-th];\\
  B.~Assel, K.~Christodoulides, A.~E.~Faraggi, C.~Kounnas and J.~Rizos,
  ``Exophobic quasi-realistic heterotic string vacua,''
  Phys.\ Lett.\  B {\bf 683} (2010) 306
  [arXiv:0910.3697 [hep-th]];\\
  A.~E.~Faraggi, C.~Kounnas and J.~Rizos,
  ``Chiral family classification of fermionic $\Z_2 \times \Z_2$ heterotic orbifold models,''
  Phys.\ Lett.\ B {\bf 648} (2007) 84
  [hep-th/0606144].
  
      \bibitem{Casimir} 
    A. Kehagias and H. Partouche, ``The Casimir effect in string theory'', CPHT-RR119.122018.


  \bibitem{CosmoPheno} 
R.~H.~Brandenberger and C.~Vafa,
  ``Superstrings in the early universe,''
  Nucl.\ Phys.\  B {\bf 316} (1989) 391;\\
G.~Veneziano,
  ``Scale factor duality for classical and quantum strings,''
  Phys.\ Lett.\  B {\bf 265} (1991) 287;\\
A.~Tseytlin and C.~Vafa,
  ``Elements of string cosmology,''
  Nucl.\ Phys.\  B {\bf 372} (1992) 443
  [arXiv:hep-th/9109048];\\
  M.~Gasperini and G.~Veneziano,
  ``Pre-big bang in string cosmology,''
  Astropart.\ Phys.\  {\bf 1} (1993) 317
  [arXiv:hep-th/9211021];\\
M.~Gasperini, M.~Maggiore and G.~Veneziano,
  ``Towards a nonsingular pre - big bang cosmology,''
  Nucl.\ Phys.\  B {\bf 494} (1997) 315
  [arXiv:hep-th/9611039];\\
  M.~Gasperini and G.~Veneziano,
  ``Singularity and exit problems in two-dimensional string cosmology,''
  Phys.\ Lett.\  B {\bf 387} (1996) 715
  [arXiv:hep-th/9607126];\\
R.~Brustein, M.~Gasperini and G.~Veneziano,
  ``Duality in cosmological perturbation theory,''
  Phys.\ Lett.\  B {\bf 431} (1998) 277
  [arXiv:hep-th/9803018];\\
  M.~Gasperini and G.~Veneziano,
  ``The pre - big bang scenario in string cosmology,''
  Phys.\ Rept.\  {\bf 373} (2003) 1
  [arXiv:hep-th/0207130].
  



\end{thebibliography}
\end{document}